\documentclass[a4paper,fleqn]{cas-dc}

\usepackage[utf8]{inputenc}
\usepackage{amsmath}
\usepackage{amsthm}
\usepackage{amssymb}
\usepackage{bm}
\usepackage{graphicx}
\usepackage{longtable}
\usepackage{booktabs}
\usepackage{circuitikz}
\usepackage{svg}
\usepackage{pdfpages}
\usepackage{float}
\usepackage[inline,shortlabels]{enumitem}
\usepackage{stfloats}
\usepackage{supertabular}
\usepackage{booktabs}
\usepackage{caption}
\usepackage[numbers, sort&compress]{natbib}
\usepackage{flushend}
\usepackage{tikz}
\usepackage{eurosym} % nicer Euro symbol, IMHO
\DeclareUnicodeCharacter{20AC}{\euro}
\usepackage{mathtools}

\newlist{E}{enumerate}{1}
\setlist[E]{label=\textbf{E\arabic*:}}

\usepackage[nameinlink]{cleveref}
\crefname{figure}{Fig.}{Figs.}
\Crefname{figure}{Figure}{Figures}
\crefname{equation}{Eq.}{Eqs.}
\Crefname{equation}{Equation}{Equations}
\crefname{section}{Section}{Sections}
\crefname{table}{Table}{Tables}
\crefname{appendix}{Appendix}{Appendices}

\newcommand{\eg}{e.g.,\ }

\newcommand{\ie}{i.e.,\ }

\DeclareMathOperator*{\argmax}{\arg \max}

\def\MIPrice{\lambda^{\text{incr}}}     % Marginal Incrmental Price
\def\MDPrice{\lambda^{\text{decr}}}     % Marginal Decremental Price
\def\tq{{t^{\circ}_{\text{q}}}}         % Start of quarter hour
\def\tQ{{t^{\dagger}_{\text{q}}}}       % End of quarter hour 
\def\alphaImb{\alpha_{\text{imb}}}      % Addend in the imbalance formula
\def\SI{\text{SI}}                      % Instantaneous System Imbalance
\def\avgSI{\tilde{\SI}}                 % Averaged System Imbalance
\def\NRV{\text{NRV}}                    % Instantaneous NRV
\def\priceFormula{\tilde{\lambda}}      % Price obtained by the formula
\def\pricePublished{\lambda}            % Published price
\def\responseInaccuracy{\gamma_{\text{resp}}}

\begin{document}

\thinmuskip=1\thinmuskip
\medmuskip=1\medmuskip
\thickmuskip=1\thickmuskip

\title[mode = title]{Predicting and Publishing Accurate Imbalance Prices Using Monte Carlo Tree Search}
\shorttitle{Predicting and Publishing Accurate Imbalance Prices Using MCTS}

\author[1]{Fabio Pavirani}[orcid=0009-0005-7904-099X]
\ead{fabio.pavirani@ugent.be}
\credit{Conceptualization, Methodology, Software, Formal analysis, Investigation, Visualization, Writing - Original Draft}

\author[1]{Jonas Van~Gompel}[orcid=0000-0002-4253-5842]
\credit{Software, Formal analysis, Writing - Review \& Editing}

\author[1]{Seyed Soroush Karimi~Madahi}[orcid=0000-0001-8072-4532]
\credit{Software, Writing - Review \& Editing}

\author[2]{Bert Claessens}[orcid=0009-0006-6116-1483]
\credit{Supervision, Conceptualization, Writing - Review \& Editing}

\author[1]{Chris Develder}[orcid=0000-0003-2707-4176]
\credit{Supervision, Funding acquisition,
Writing - Review \& Editing}

\affiliation[1]{organization={IDLab Ghent university -- imec},
                addressline={Technologiepark Zwijnaarde 126}, 
                postcode={9052}, 
                postcodesep={}, 
                city={Gent},
                country={Belgium}}
\affiliation[2]{organization={Beebop},
                country={Belgium}}
                
\date{April 2024}

% Short author
\shortauthors{Pavirani et al.}

\bibliographystyle{cas-model2-names}

% Keywords
% Each keyword is separated by \sep

\graphicspath{{pictures}}

\makeatletter\def\Hy@Warning#1{}\makeatother
\maketitle

%================================================
\begin{abstract}
%================================================
The growing reliance on renewable energy sources, particularly solar and wind, has introduced challenges due to their uncontrollable production. This complicates maintaining the electrical grid balance, prompting some transmission system operators in Western Europe to implement imbalance tariffs that penalize unsustainable power deviations. These tariffs create an implicit demand response framework to mitigate grid instability. 
Yet, several challenges limit active participation. In Belgium, for example, imbalance prices are only calculated at the end of each 15-minute settlement period, creating high risk due to price uncertainty. This risk is further amplified by the inherent volatility of imbalance prices, discouraging participation. Although transmission system operators provide minute-based price predictions, the system imbalance volatility makes accurate price predictions challenging to obtain and requires sophisticated techniques. Moreover, publishing price estimates can prompt participants to adjust their schedules, potentially affecting the system balance and the final price, adding further complexity.
To address these challenges, we propose a Monte Carlo Tree Search method that publishes accurate imbalance prices while accounting for potential response actions. Our approach models the system dynamics using a neural network forecaster and a cluster of virtual batteries controlled by reinforcement learning agents. 
Compared to Belgium’s current publication method, our technique improves price accuracy by 20.4\% under ideal conditions and by 12.8\% in more realistic scenarios. This research addresses an unexplored, yet crucial problem, positioning this paper as a pioneering work in analyzing the potential of more advanced imbalance price publishing techniques.
\end{abstract}

%================================================
\section{Introduction}
\label{sec:intro}
%================================================
% Problem 1
In light of recent developments in climate change research, decarbonization of energy sources is a necessary step to reduce global CO2 emissions~\cite{arantegui2018photovoltaics}. 
% Solution 1
For this reason, an increasing amount of Renewable Energy Sources (RES) is being installed~\cite{papadis2020challenges, arantegui2018photovoltaics, irena2022weto}. 
% Problem 2
Although RES such as photovoltaic panels and wind farms have low carbon emissions, their energy supply is hard to accurately predict due to weather dependencies. 
This complicates the task of balancing power production and consumption, with direct consequences for the grid's operational safety.
% Solution 2
Hence, the electrical grid is being gradually redesigned to handle local and uncertain variations of electrical production~\cite{rahim2022overview}. For instance, some Transmission System Operators (TSOs) in Europe offer remuneration to Balancing Service Providers (BSPs) in exchange for contracted energy transactions that help balance the grid. Moreover, a specific imbalance tariff is applied to Balance Responsible Parties (BRPs), penalizing deviations in their energy schedules that disrupt the balance of the grid. 
By strategically designing the imbalance fee structure, TSOs can create an implicit Demand Response (DR) framework for the BRPs, which will then react to the prices and help to reduce the grid's System Imbalance (SI) magnitude. 

\begin{figure}
    \centering
    \includegraphics[width=0.5\textwidth]{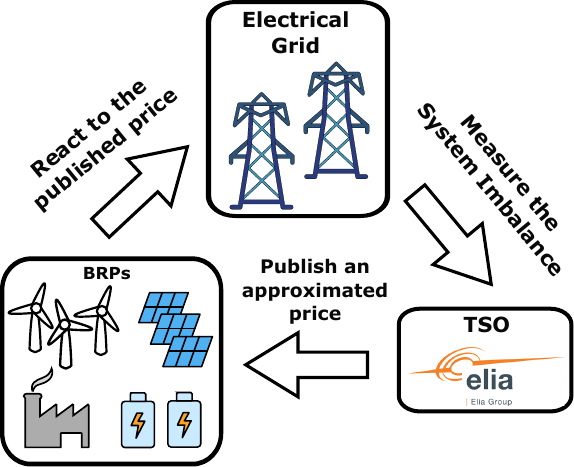}
    \caption{Influence cycle between the BRPs and the TSO. The TSO publishes an approximation of the imbalance price, triggering an implicit reaction from the BRPs. The reaction influences the grid's SI, hence impacting the final imbalance price at the closure of the settlement period.}
    \label{fig:influence_cycle}
\end{figure}

% Problem 3
In Belgium, the imbalance prices are calculated at the \emph{end} of each settlement period, spanning 15 minutes. This implies that BRPs face significant risk, as each transaction is subject to a price that remains unknown at the time of execution.
This obstructs active participation from BRPs in the imbalance mechanism, as the prices might end up being less attractive than the expected ones, or even unprofitable for BRPs in case a significant swing of the SI occurs within the quarter hour.
% Solution 3
To tackle this, the Belgian TSO publishes an estimation of the imbalance price \emph{within} each quarter hour on a minute basis, reducing the risks involved in the imbalance settlement participation.
% Problem 4
However, the imbalance prices are dependent on the grid's SI, making these values highly volatile. Hence, obtaining an accurate approximation of the imbalance prices is challenging. Moreover, publishing real-time prices might trigger an implicit reaction from BRPs, altering the SI and thereby impacting the price determined at the closure of the settlement period. By implicit response, we mean a voluntary, though undeclared, deviation of BRPs energy schedules aimed at taking advantage of the imbalance prices.
This recursive cycle (illustrated in \cref{fig:influence_cycle}) adds complexity to the problem, highlighting a significant potential for the application of advanced planning techniques in the price publication problem. 
The increasing implicit participation in the Belgian imbalance settlement, together with the lack of proper methods for publishing approximated prices, motivates us to tackle this problem with a more advanced technique compared to the currently used one.

% 'Solution' 4
To address the challenge of publishing accurate real-time imbalance prices, two main approaches can be chosen: a model-based, and a model-free one.
A \emph{model-based} approach uses a model of the system's dynamics to obtain each prediction (and publication) of the final price. Conversely, a \emph{model-free} approach would obtain predictions by trying to anticipate the future behavior of the system without directly formulating a model of the latter. 
Given the complexity of the task, we believe that the most promising approach is model-based, allowing the publication technique to access insightful information about the grid's dynamics, thereby enhancing the accuracy of price predictions.

% Problem 5
To solve the problem, a good technique needs to:
\begin{enumerate*}[\bfseries (a)]
    \item provide accurate publications of the imbalance price,\label{it:accuracy}
    \item have a low computational time for each prediction to enable real-time (minute-based) price publications, and\label{it:time}
    \item % allow flexibility regarding the complexity of the models involved in the prediction, \ie the technique must be able to handle complicated models to perform its predictions. 
    offer flexibility in handling varying levels of model complexity, enabling it to accommodate intricate models in its predictions.\label{it:flexibility} 
\end{enumerate*}
We propose a solution based on Monte Carlo Tree Search (MCTS), which is a search technique that mostly gained importance in board games such as Go and Chess~\cite{silver2017mastering,silver2018general,schrittwieser2020mastering}. 
MCTS has reached state-of-the-art performance in several complex decision making problems, some of those requiring fast computations due to time limitations, effectively addressing \cref{it:accuracy,it:time}. 
Moreover, the only requirement for MCTS is to have a simulable environment that approximates the dynamics of the system. This allows for highly non-linear models such as Neural Networks (NNs) based forecasters, addressing \cref{it:flexibility}. 

To model the system's stochastic dynamics, we deployed an NN-based forecaster of the SI with a prediction horizon of 15 minutes and a granularity of 1 minute. The forecaster is composed of an ensemble of NNs, based on~\citet{van2024probabilistic}, using a Constant Variable Selection Network (C-VSN) architecture to accurately forecast the SI. 
We then modeled the BRPs implicit reaction to the published imbalance prices using a cluster of virtual batteries. Each battery is controlled by a Reinforcement Learning (RL) agent, similar to~\cite{madahi2023distributional}. 
By doing so, our publication technique takes into account the effect of its publication on the grid's balance, enhancing its prediction accuracy. 
% The joint output of these structures (NN-based forecaster and cluster of deep RL agents) creates the model used by MCTS to roll out different action trajectories.
% To analyze the importance of the batteries' capacity, we considered different battery sizes, each representing a different degree of influence the batteries exert on the grid. 
% Overall, our publication technique uses different machine learning algorithms to generate its prediction. 
In summary, our approach adopts 
\begin{enumerate*}[(i)]
    \item 
    \label{it:system_model}
    a system model to predict SI evolution, which is used in 
    \item 
    \label{it:publication_model}
    a model-based RL algorithm to generate the prices to publish.
\end{enumerate*}
For \cref{it:system_model}, a \textit{model of the system} is obtained using the joint outputs of an advanced NN ensemble forecaster and a cluster of model-free RL agents. For \cref{it:publication_model}, we use MCTS.
% The model is then used by a model-based RL algorithm -- MCTS -- to generate valuable imbalance price publications. 

Using Belgian data from 2023, we quantitatively analyze our technique in terms of such as price prediction accuracy, influence on the grid's SI, TSO balancing costs, and practical feasibility. Our contributions can be summarized as follows:
\begin{enumerate}
    \item \textbf{We propose a planning technique for publishing prices:} We deployed and analyzed an MCTS technique for a TSO to predict and publish accurate imbalance prices. Through simulations, the technique is benchmarked with the current publication method used in Belgium, obtaining superior results in terms of price accuracy.
    \item \textbf{We assess the impact of such technique on the grid:} We analyzed the effect of the proposed technique in terms of grid stability. We tracked grid-related values such as SI and balancing costs, and benchmarked them with the ones obtained by the baseline. Moreover, we modify the objective function to investigate the technique's potential when incorporating additional balancing objectives beyond price accuracy. The results show the positive impact of using our technique on grid stability.
    \item \textbf{We assess the practical applicability of the technique:} Finally, we evaluated the technique's feasibility by integrating an SI forecaster\footnote{Later called as Net Regulation Volume (NRV) forecaster} and implicit response inaccuracies in the system dynamics model; thus showing the applicability of the method even when dealing with uncertain conditions. 
\end{enumerate}
Compared to the current publication method used in Belgium, our technique provides a price accuracy improvement of up to 20.4\% 
when provided with ideal conditions (\ie perfect knowledge of the grid dynamics) and up to 12.8\% when considering more realistic conditions (\ie forecasted SI and uncertain BRPs response).  

To the best of our knowledge, this is the first time an MCTS technique has been proposed as part of a pricing method in a DR scenario. Moreover, this is to our understanding the first research work that predicts and publishes real-time imbalance prices from a TSO point of view.

%================================================
\section{Related Work} %Literature review}
\label{sec:literature_review}
%================================================

Over the last decades, an increasing number of studies have focused on energy DR, which implies adapting the energy \emph{demand} to make it better match the available supply, in contrast to
conventional techniques that only shape the energy production. This usually happens by sending a signal that reflects the supply status (\eg a price signal) to the demand assets~\cite{shen2014role}. The demand components are thus prompted to react to the signal, shaping their consumption.
The implementation of DR programs significantly helps decarbonization of the energy transition~\cite{chantzis2023potential}, hence highlighting the relevance of the topic.

The majority of previous studies focused on how the demand side can react to such signals to optimize revenues.
In this context, RL techniques have demonstrated considerable promise~\cite{vazquez2019reinforcement}. 
For example, \cite{jiang2021building} developed a Deep-RL solution to reduce the cost of an HVAC unit subjected to dynamic prices. 
RL has also been deployed in problems requiring multiple assets or structures to work parallelly. For example, the authors of~\cite{xie2023multi,ajagekar2024energy} implemented a Multi-Agent Deep Reinforcement Learning framework to perform energy management in a DR mechanism applied to residential or industrial contexts. 
Moreover, following the need for artificial intelligence methods to provide trustworthy control actions, \citet{yun2023explainable} implemented a decentralized multi-agent RL framework with enhanced explainability in an industrial setting. 
These studies are just an example of the relatively recent growing interest in RL algorithms
to realize effective DR control.

Despite the extensive amount of studies regarding DR, the majority of them focus on creating an effective reaction of the demand side. Instead, only a small portion of them focused on how to design the DR framework from an energy supplier perspective. \citet{theate2023matching} investigated the problem of designing a residential dynamic pricing system through mathematical definitions of the decision-making problem. Moreover, they presented a discussion over the necessary algorithms necessary to generate price signals that maximize the synchronization between the demand side and the supply side. \citet{lu2018dynamic} formulated the problem using a Markovian Decision Process (MDP) and approached it using a Q-learning algorithm. \citet{salazar2023reinforcement} also used an RL technique to design a DR framework with both incentive- and price-based signals. Works like~\cite{jia2016dynamic,meng2013stackelberg} modeled the dynamic pricing problem using Stackelberg games to obtain theoretically optimal results for both energy retailers and customers. Finally, \citet{lai2022multiagent} modeled the pricing problem as a Stackelberg game and solved it using RL, until a near Stackelberg equilibrium was reached. %\\
In general, even though a clear effort has been made to cover the pricing topic, more analyses and investigations should be produced. Specifically for applying RL to design a DR framework, the previously cited works only considered simple algorithms such as Q-learning. More advanced techniques should be analyzed.

Regarding the imbalance settlement mechanism, several works have studied how to effectively react to the received prices. In particular: given the significant risks imbalance participants have to face, there is a need for policies that take these sensitivities into account.
\citet{smets2023strategic} implemented a risk-sensitive policy for a battery energy storage system (BESS) in imbalance settlement using a stochastic optimization technique. They forecasted the future SI with an attention-based recurrent neural network and optimized the participation of a BESS using a MILP technique. 
Similarly, \citet{madahi2023distributional} implemented a risk-sensitive policy paired with a BESS to leverage the imbalance prices. However, differently from \cite{smets2023strategic}, they used a distributional RL algorithm to add a risk-appetite parameter in defining the control objective.
Because of the high volatility of the imbalance prices, work such as~\cite{smets2023strategic,narajewski2022probabilistic,dumas2019probabilistic,ganesh2023forecasting,bottieau2019very} focused on predicting the next imbalance prices. However, different from our work, these studies were performed from a BRP perspective, and their goal was to further optimize the demand side profits.

Thus, to the best of our knowledge, no previous work has focused on predicting (and publishing) the imbalance prices from a TSO perspective.
Overall, we identify the exploitation of the imbalance settlement as a not yet broadly explored topic in literature. Moreover, the specific task for TSOs of publishing accurate prices before the closure of each settlement period is, to our knowledge, mostly unexplored.
We hence propose an MCTS-based technique to predict and publish prices that accurately approximate the actual one obtained at the closure of the settlement period. It is important to note that, although we identify the literature on DR pricing techniques as the closest to our work, we do not propose a pricing method that defines how to design the imbalance tariffs, but rather we focus on how to obtain the published value that best approximates the final price charged to BRPs. 

%===============================
\section{Problem Description}
\label{sec:problem_formulation}
%===============================

%-------------------------------
\subsection{Balancing services}
\label{sec:balance_services}
%-------------------------------
TSOs are responsible for the correct functioning of the electrical grid. For this reason, with the increasing amount of RES affecting the grid's stability, TSOs have become more dependent on ancillary services that are designed to restore and maintain the grid's frequency balance.
In Western Europe, three distinct services are active, each requiring different reaction times to frequency deviations and system imbalances. 
In order of decreasing reaction speed, these services are respectively: Frequency Containment Reserve (FCR/R1), automatic Frequency Restoration Reserve (aFRR/R2), and manual Frequency Restoration Reserve (mFRR/R3). We are particularly interested in the secondary (aFRR) and tertiary (mFRR) reserves, as their activations directly influence the imbalance price determination in each settlement period (see \cref{sec:imbalance_pricing} for details). 

Participation in aFRR is provided through two different regulations: incremental and decremental. The activation of these two regulations depends on the sign of the system imbalance (SI). A positive SI (surplus of power) activates the decremental regulation, asking the BSPs to withdraw power from the grid (\ie consume more or produce less); and a negative SI (lack of power) activates the incremental regulation, asking the BSPs to inject power into the grid (\ie consume less or produce more). For each activation period (15 minutes), each regulation receives a list of bids from the BSPs. Each bid contains a price [\euro/MWh] and a volume [MW]. The bids get sorted by the TSO based on their economic value, \ie the TSO identifies which bids would minimize their cost and sorts them accordingly. 
\begin{figure}
    \centering
    \includegraphics[width=0.5\textwidth]{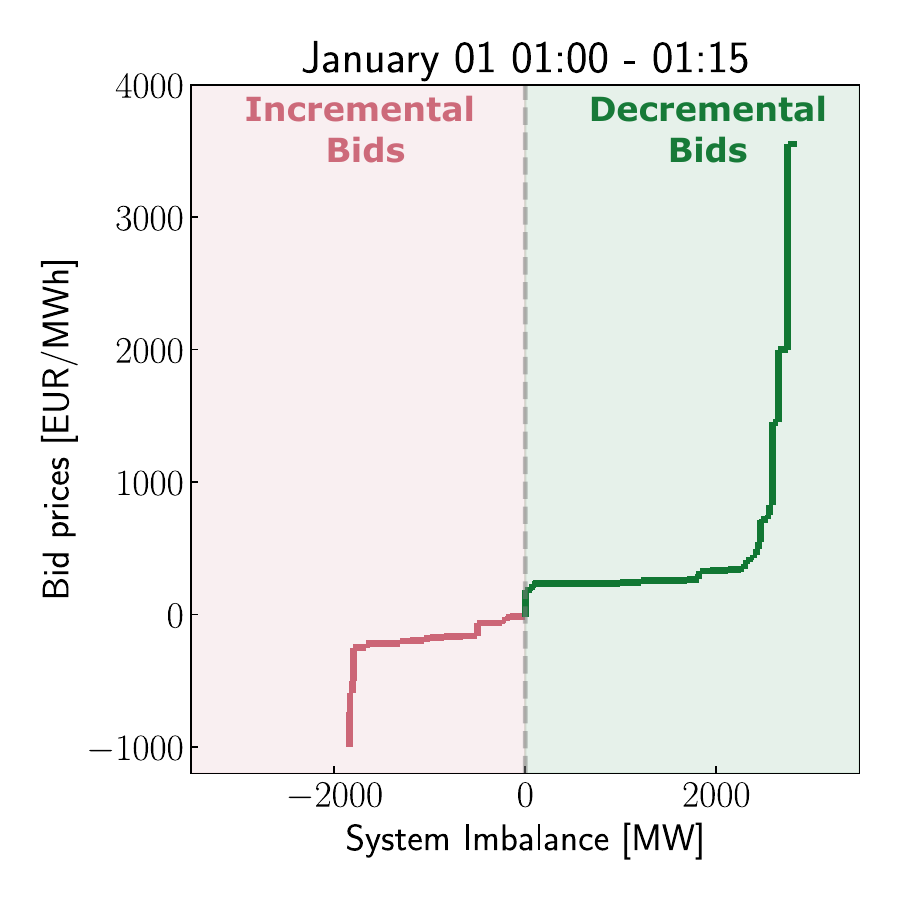}
    \caption{Sample bid ladder, generated using Belgian data from 2023.}
    \label{fig:bid_ladder_example}
\end{figure}
\begin{figure*}
    \centering
    \includegraphics[width=1\textwidth]{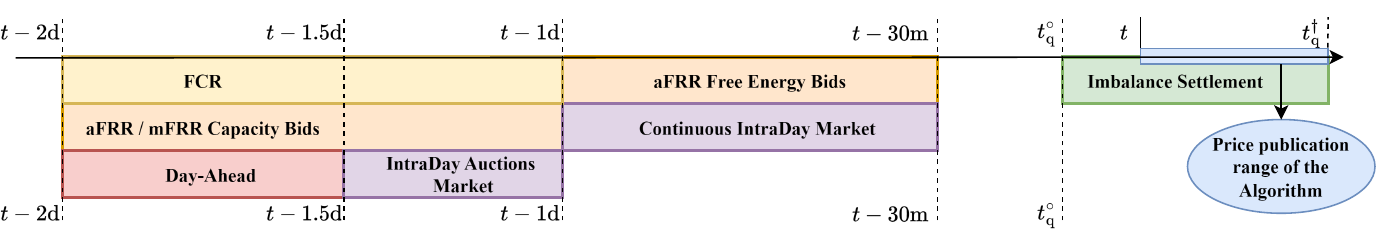}
    \caption{Timeline of the major energy markets and services, and the time position of where our algorithm would work.}
    \label{fig:energy_timeline}
\end{figure*}
\begin{figure}
    \centering
    \includegraphics[width=0.5\textwidth]{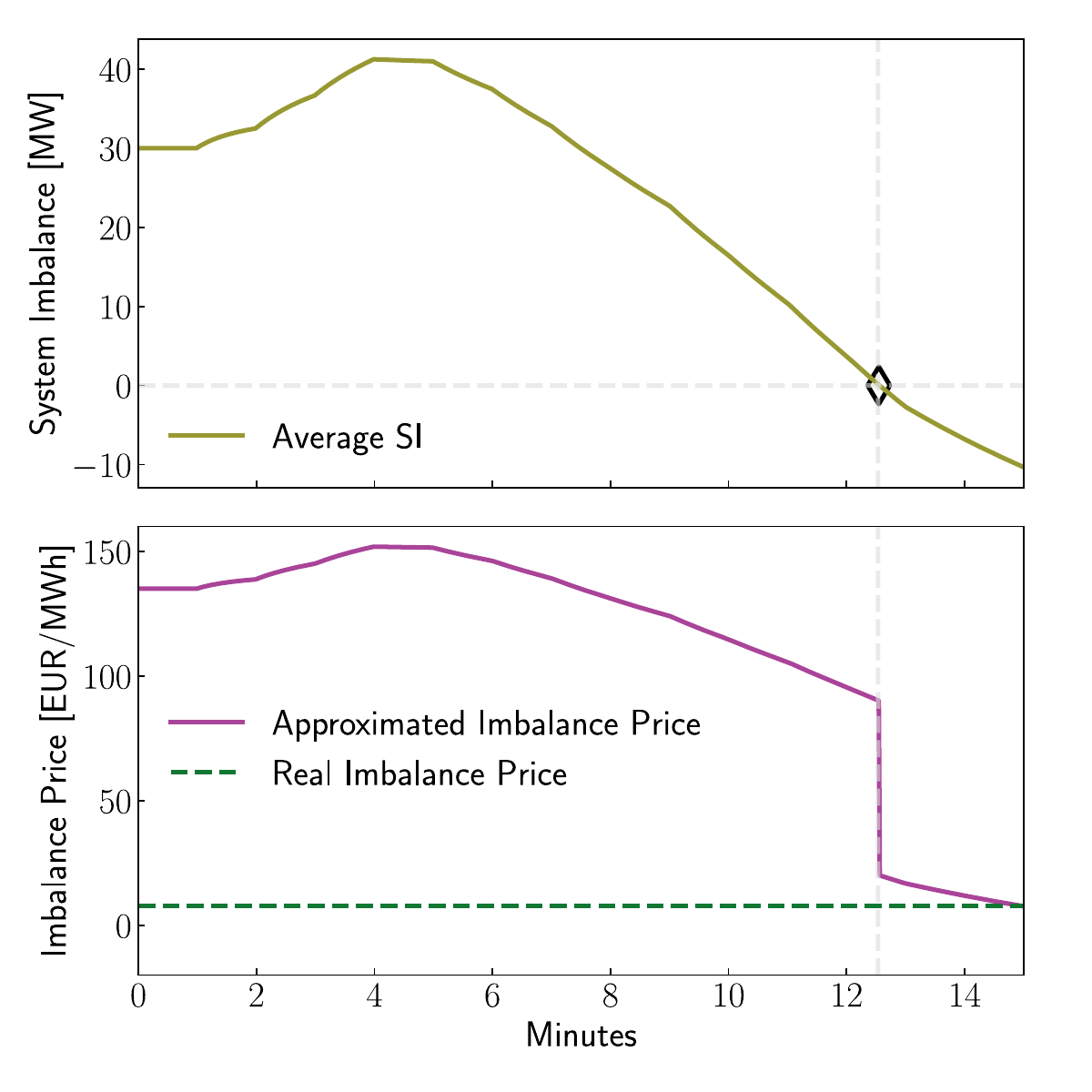}
    \caption{Example showing a possible progression of the imbalance prices following a change in sign of the system imbalance. The imbalance price formula applied every minute of the quarter hour does not accurately predict the final price when a change in sign of the system imbalance occurs close to the end of the quarter.}
    \label{fig:change_sign_si}
\end{figure}
The sorted bids for each regulation create a \emph{bid ladder} that, depending on the SI of the activation period, will determine the imbalance price. 

mFRR uses an activation mechanism akin to that of aFRR, the main difference being that mFRR requires a less strict reaction time and activation happens only after the aFRR capacity has been fully addressed. An example of a bid ladder is given in \cref{fig:bid_ladder_example}.

Given the merged bid ladders of aFRR and mFRR, and given the current SI (and, consequently, the activated balancing volume), the marginal incremental price $\MIPrice_t$ in a certain timestep $t$ is defined as the last activated bid in the incremental merged bid ladder (\ie the most expensive activated bid for the TSO). Similarly, the marginal decremental price $\MDPrice_t$ is the last activated bid in the decremental merged bid ladder.

%------------------------------------------------------------
\subsection{Imbalance tariff and pricing system}
\label{sec:imbalance_pricing}
%------------------------------------------------------------

%----------------------------------------------------
\subsubsection{Belgian Imbalance Price Formula}
%----------------------------------------------------
As previously stated in \cref{sec:intro}, some European TSOs designed a specific tariff addressed to BRPs that deviate from their previously declared production/consumption values. This tariff, also called \textit{imbalance settlement mechanism}, is not standardized in Europe, \ie each TSO is free to decide how to design the pricing formula. However, an electricity balancing guideline (EBGL) has been published by the European Network of Transmission System Operators for Electricity (ENTSO-E).
According to the EBGL: \textit{``The general objective of imbalance settlement is to ensure that Balance Responsible Parties support the system balance in an efficient way, and to incentivize market participants in keeping and/or helping to restore the system balance [\ldots]''}~\cite{european_regulation}. 
The pricing formulas can then be designed to promote grid stability through mechanisms aligned with open-market principles.
Moreover, the settlement `transactions' can be done in a real-time framework (as observable in \cref{fig:energy_timeline}), hence creating a desirable tool for BRPs that wish to constantly update and correct their energy positions. 

In Belgium, the imbalance prices are determined based on the marginal balancing activated prices, and the periodicity of imbalance settlement in Belgium is 15 minutes. 
We use $t \in \mathbb{N}$ as a timestep index, each spanning 1 minute. We refer to $\tq \in \mathbb{N}$ as the timestep index corresponding to the start of the quarter hour $t$ belongs to, and $\tQ \in \mathbb{N}$ as the timestep index corresponding to the end of the quarter hour $t$ belongs to. In other words, given a (minute-based) timestep $t$, then ${\tq \doteq t - (t \mod 15) \leq t}$ and {$\tQ \doteq \tq + 14 \geq t$}.
% $\tQ \doteq t + 15 - (t \mod 15) \geq t$. 

The price formula used in Belgium is then:\footnote{Please note that Elia adjusted the price calculation on 20/07/2024. To maintain consistency with historical data and because the change is not particularly significant, we will continue using the previous formula for our experiments.}
\begin{equation}
\label{eq:price_formula}
    \priceFormula_t \doteq 
    \begin{cases}
        \MDPrice_t - {\alphaImb}_{,\,t} & \text{if: } \avgSI_t > 0\\
        \MIPrice_t + {\alphaImb}_{,\,t} & \text{if: } \avgSI_t < 0 
    \end{cases}
\end{equation}
where ${\alphaImb}_{,\,t}$ is a correction addend used to further incentivize imbalance participation in cases where the SI's magnitude is particularly high, and $\avgSI_t$ is the averaged System Imbalance measured since $\tq$. 
\begin{equation}
\label{eq:average_SI}
    \avgSI_t \doteq \frac{1}{t-\tq}\sum_{t=\tq}^{t}{\SI_t}
\end{equation}
Details about the calculation of ${\alphaImb}_{,\,t}$ can be found in \cref{app:alpha}.
Note that the actual price that will be charged to BRPs gets calculated \emph{at the end} of the quarter hour, \ie at timestep $\tQ$. However, the formula is generic and can be used for every minute-based timesteps to get an approximation of the final price.

Given this formula, because of the balancing structure discussed above, we can then expect a simple yet effective relation between the grid's SI and the imbalance price, where a high SI would lead to low imbalance prices (coming from the decremental regulation), and a low SI would conversely lead to high imbalance prices (coming from the incremental regulation). BRPs that deviate from their declared energy schedules by helping balance the grid will hence obtain economic remuneration, as they will end up leveraging the prices. Conversely, deviations that hinder the grid balance will result in non-profitable costs for the BRPs.
Thus, profitable participation by BRPs in the imbalance settlement mechanism can assist the TSOs in maintaining the SI within operational limits.

%------------------------------------------------------------------
\subsubsection{Belgian Imbalance Settlement}
\begin{figure*}
    \centering
    \includegraphics[width=1\textwidth]{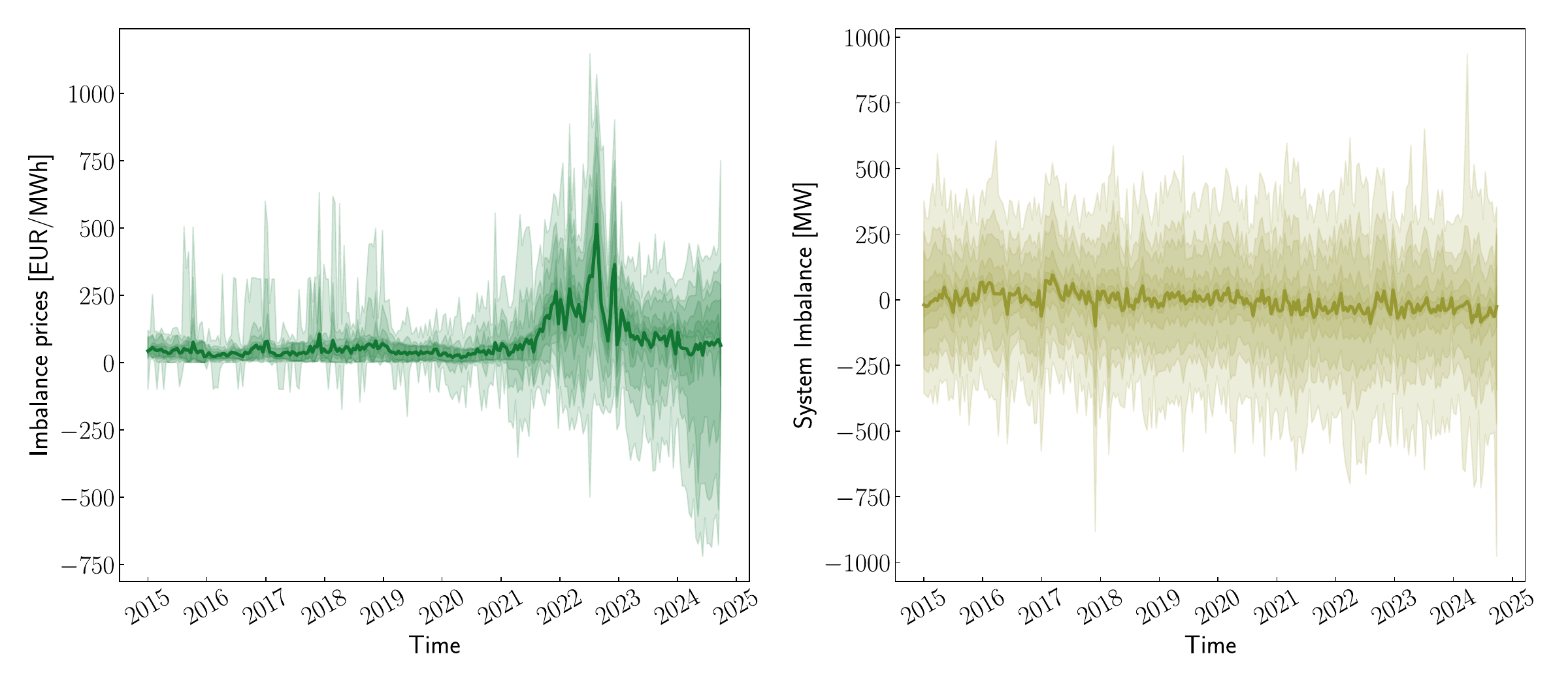}
    \caption{Historical analysis of the applied imbalance prices (left) and of the system imbalances (right) in Belgium, showing the mean value (line) and different quantile values (colored bands) up to the 1\%-99\% quantile interval. In the last few years, the imbalance prices have been facing a remarkable increase in magnitude. This seems to be independent of the SI deviations, which have been mostly stable for the last decade. This increment in imbalance price magnitudes creates a big opportunity for BRPs that wish to exploit the mechanism. As this trend continues, we can then expect the imbalance implicit responses to grow.}
    \label{fig:historical_published_price_analysis}
\end{figure*}

\begin{figure*}
    \centering
    \includegraphics[width=1\textwidth]{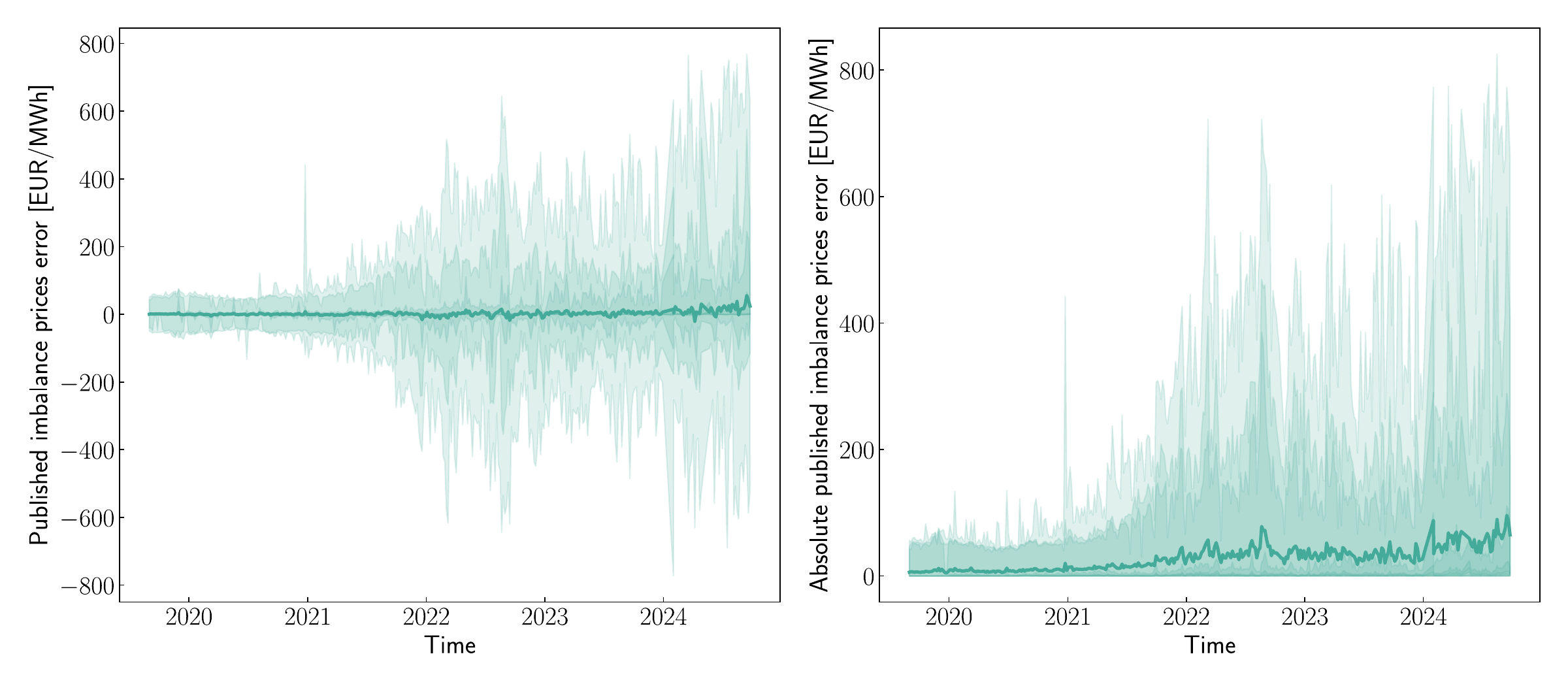}
    \caption{Visualization of the historical imbalance prices publication error (left)
    and absolute error (right).
    The lines correspond to the mean value, while the colored bands correspond to different quantile values, up to the 1\%-99\% quantile interval. The data, downloaded from the Belgian TSO, contains some missing values. In that case, we linearly interpolated the data holes. We observe a recent increase in publication errors, highlighting the need for more accurate publication techniques.}
    \label{fig:historical_published_prices_error}
\end{figure*}

%------------------------------------------------------------------
Despite the general guidelines provided by the {ENTSO-E}, TSOs in Europe have designed pricing systems with objectives that can substantially vary from country to country. For instance, some countries designed the imbalance settlement as a tariff mechanism to incentivize the BRPs to \emph{always} keep their balance, penalizing implicit deviations no matter the sign of the SI. Conversely, other countries -- such as Belgium -- consider the imbalance settlement as a promising tool to exploit BRPs' flexible assets to implicitly balance the grid, in a process also referred to as passive balancing. Therefore, it is in the Belgian TSO's interest to obtain a more active BRP engagement in the imbalance settlement.
Despite the potential benefits of active participation in the imbalance settlement, a direct engagement of BRPs contains non-negligible risks. This mainly stems from the uncertain and volatile nature of the grid's SI, and hence of the imbalance prices. Moreover, the prices are only fixed at the end of the settlement period. Participating in the imbalance settlement hence implies making energy transactions at an unknown price, leading to substantial risks for BRPs. 
In an attempt to alleviate the risks discussed above (and, consequently, trigger stronger participation from BRPs), the Belgian TSO publishes an approximation of the final imbalance price at a minute pace \emph{within} the settlement period. This is currently done by sending the value obtained using \cref{eq:price_formula} every minute in each settlement period. While this approximation provides an important reduction of the risk BRPs are forced to face, it might still fail to provide accurate predictions of the final price. For example, a change of sign of the SI during the last minutes of the quarter hour can generate a drastic change of the final imbalance price that will not be predicted by the formula in \cref{eq:price_formula}. This phenomenon is better illustrated in \cref{fig:change_sign_si}. 
Moreover, the formula does not take into account the reaction that BRPs might have in response to the real-time published prices. With an increasing magnitude of the imbalance prices (as shown in \cref{fig:historical_published_price_analysis}), we can expect a growth of BRPs responses. If we do not explicitly consider these responses in the prediction model, we can expect the publication results to be less and less accurate as the response magnitude increases. We will delve into this problem in \cref{sec:results_exp1.5}.
\Cref{fig:historical_published_prices_error} shows the publication error in Belgium using historical data from 2019. We observe that publication errors increased significantly in the last few years, highlighting the need for a more accurate publication technique.
For these reasons, we believe using more advanced techniques can help both the TSO %s
in assuring the operational functioning of the grid, and the BRPs to effectively exploit the imbalance settlement. 

Note that in the next sections, we will mainly consider the Net Regulation Volume (NRV). The NRV is defined as the difference between the gross incremental regulation volume and the gross decremental regulation volume. For this reason, it is highly correlated to the SI (specifically: $\NRV\,\approx\,{-\SI}$). We will then approximate the grid's SI as $-\NRV$.

%===========================================
\section{Methodology}
\label{sec:methods}
%===========================================

%-----------------------------
\subsection{NRV forecaster}
\label{sec:forecaster}
\begin{figure*}
    \centering
    \includegraphics[width=1\textwidth]{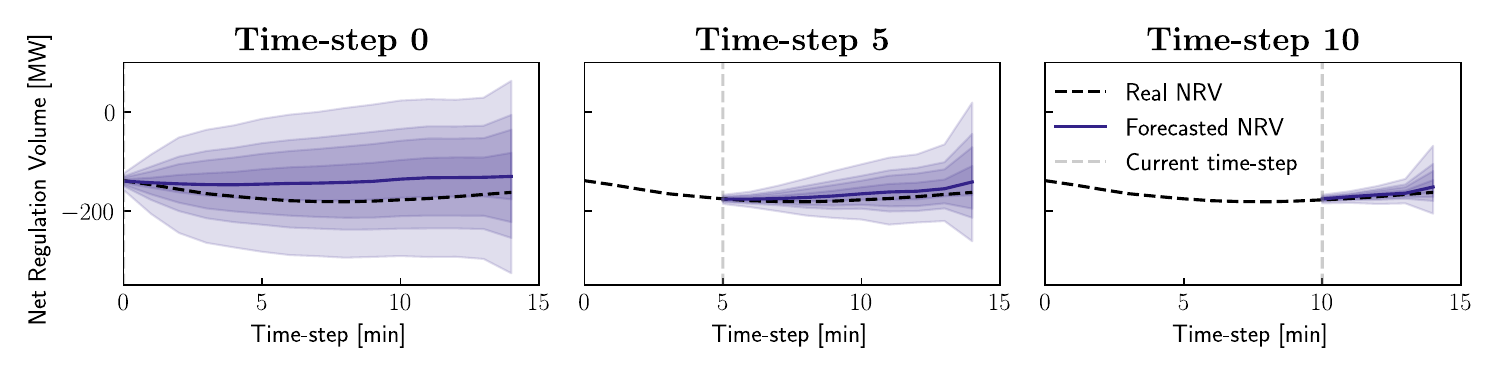}
    \caption{Example of output obtained from the NRV forecaster in three different timesteps of a quarter hour. When integrated with the MCTS technique, the forecaster is used every minute of each quarter hour. Thanks to the structure of the model~\cite{van2024probabilistic}, the forecaster provides probabilistic predictions. However, in this work, we only use deterministic predictions (\ie the full line in the graphs).}
    \label{fig:forecaster_plot}
\end{figure*}

%-----------------------------
To predict the NRV for future timesteps, we use a specific neural network architecture called constant variable selection networks (C-VSN), similar to what was proposed in~\cite{van2024probabilistic}. The forecaster was originally built to predict the SI. However, by modeling the NRV as the SI's inverse ($\NRV \sim -\SI$), we can easily adapt the model to our needs.
The model consists of an ensemble of 21 neural networks, each differing from the others by using various training seeds, bootstrapping techniques, output formats, and loss weights.
As explained in the original work~\cite{van2024probabilistic}, the model is trained by adding a variable weight to the training loss function, specifically aimed to increase the accuracy performance during SI spikes. This feature, together with the overall accuracy performance obtained, makes the model appropriate for the needs of our price publication problem. 
The main differences from the model described in~\cite{van2024probabilistic} are:
\begin{itemize}[noitemsep,topsep=0pt]
    \item Our model outputs the predicted minute-based NRVs for the next 15 minutes, instead of the quarter hour averaged SI up to 45 minutes ahead of time.
    \item To obtain higher independence of the model predictions from the implicit responses to published imbalance prices, we removed the input features that are mostly correlated to that (\eg the imbalance prices input values).
\end{itemize}
From the input features used~\cite{van2024probabilistic}, we only include 
\begin{itemize}[noitemsep,topsep=0pt]
    \item Time features,
    \item SI and NRV history,
    \item Net cross-border nominations,
    \item PV and Wind generation forecasts, and
    \item Load forecasts.
\end{itemize}
Using this forecaster allows us to simulate the future behavior of the NRV and helps predict which imbalance price will be applied at the end of the quarter hour. An example of the forecaster outputs is shown in \cref{fig:forecaster_plot}. More details about integrating the NRV forecaster into our price publication solution are in \cref{sec:MDP_def}.

%-----------------------------------------
\subsection{Implicit Imbalance Response}
\label{sec:implicit_response}
%-----------------------------------------

While publishing the approximations of the Imbalance price of the current quarter hour, the TSO might trigger an implicit reaction from BRPs that wish to exploit the price. This reaction will influence the SI of the grid. Hence, in cases where the implicit reaction magnitude is significant, the publication of a price might change the final price itself. 
Given the recent increment in the magnitude of the BRPs' flexibility assets, we can expect such phenomena to effectively occur in the price publication problem. Therefore, it is important to independently model the implicit reaction to better capture the NRV dynamics of the system.

\subsubsection{Model structure and motivations}
We model the implicit reaction as a cluster of virtual batteries leveraging the prices. Each battery is controlled by an independent policy-based RL agent (trained offline using historical data) whose objective is to maximize the profit of the battery by participating in the imbalance settlement. The motivations for this choice can be summarized as:
\begin{itemize}
    \item Batteries are increasingly being deployed by BRPs at grid scale.
    \item Many flexible assets can be modeled as an energy buffer, which conceptually works as a battery~\cite{ulbig2015analyzing,lahariya2021physics}. %%%CHRIS%%% Hence, using a cluster of batteries can be seen as a sensible generalization of the actual reaction.
    \item Since implicit reactions are virtually impossible to measure from the publicly available data, we use a plausible and sensible model that does not require special data to assess the response of BRPs.
\end{itemize}
Last, it is worth mentioning that modeling the BRPs' reaction to published prices is not straightforward and requires particular attention to all the major asset types used in the imbalance settlement.
Moreover, even with scrupulous effort, a certain degree of uncertainty has to be expected. 
In this work, we do not claim to provide an accurate model of such complex dynamics. Instead, we limit ourselves to providing a \emph{sensible} model that enables us to validate the potential of our proposed MCTS-based technique.
Furthermore, to analyze the effect of inaccuracies in the response model, we artificially added noise in the response model to evaluate their effect on the algorithm performances (more details in \cref{sec:results_exp3}).
Next, we concisely describe the framework used to train the RL agents.

\subsubsection{Reinforcement Learning framework}
The general structure formulation of a problem that RL tries to solve is a Markov Decision Process (MDP)~\cite{busoniu2017reinforcement}. An MDP is composed by the tuple $(\mathcal{S}, \; \mathcal{A}, \; f(s, a, s'), \; \rho(s, a, s'))$, where $\mathcal{S}$ is the \emph{State Space}, $\mathcal{A}$ is the \emph{Action Space}, $f(s, a, s'):\; \mathcal{S} \times \mathcal{A} \times \mathcal{S} \rightarrow [0, 1]$ is the \emph{Transition Function}, and $\rho:\; \mathcal{S} \times \mathcal{A} \times \mathcal{S} \rightarrow \mathbb{R}$ is the \emph{Reward Function}.
Given an MDP, we can formulate an optimization problem, that is to find a sequence of actions that maximize the obtained rewards. Formally, given a \emph{policy} $\pi: \mathcal{S} \rightarrow \mathcal{A}$, we can consider the definition of its Q-function~\cite{busoniu2017reinforcement}:
\begin{equation}
    Q^\pi(s, a) \doteq \mathbb{E}_{s' \sim f(s, a, \cdot)}\left[\rho(s, a, s') + \gamma R^{\pi}(s') \right],
\end{equation}
where $\gamma \in [0, 1]$ is the discount factor, and
\begin{equation}
    R^{\pi}(s_0) \doteq \lim_{T\rightarrow +\infty}\mathbb{E}_{s_{t+1} \sim f(s_t, \pi(s), \cdot)}\left[{\sum_{t=0}^{T}{\gamma^t\rho(s_t, \pi(s_t), s_{t+1})}}\right].
\end{equation}
In other words, the Q-function $Q^\pi(s, a)$ estimates the value of acting $a$ from state $s$, and then following the policy $\pi$ afterward. Given a state $s$ and an action $a$, we can consider the optimal Q-function~\cite{busoniu2017reinforcement} as $Q^*(s,a) \doteq \max_\pi{Q^\pi(s, a)}$. The optimization problem would then be:
\begin{equation}
    \pi^*(s) \in \arg \max_a{Q^*(s, a)} \; ; \; \forall s \in \mathcal{S} \; .
\end{equation}

RL algorithms typically learn a close-to-optimal policy by repeatedly interacting with the environment and have been proven as a viable solution to deal with a high amount of uncertainty and stochasticity.
For this reason, RL algorithms received substantial attention in DR applications~\cite{vazquez2019reinforcement}. 
Specifically, given the real-time framework required, RL techniques are a promising solution to address the high volatility of the imbalance prices~\cite{lago2020optimal}. We specifically follow the work~\cite{madahi2023distributional} to model the implicit responses. 
We will consider a cluster of independent agents, each controlling a virtual battery. Each agent is based on a Soft-Actor Critic (SAC)~\cite{haarnoja2018soft,haarnoja2018soft2} previously trained on historical data to maximize the battery revenue. The agents are hard-constrained (both during training and inference) not to consume an excessive amount of daily battery cycles (specifically, the battery can only consume 1~cycle every day). That prevents the agents from providing an unrealistically high amount of energy transactions, but rather prompts the agent to learn the most profitable periods each day.

The motivations behind the choice of SAC mostly lie in the nature of the algorithm itself. Being a policy-based algorithm, it can provide continuous actions. This makes the reaction model more moderate, enabling a more realistic representation (compared to, for example, bang-bang agents that would provide more drastic fully on/off responses). 
Moreover, RL algorithms have been proven to be more effective in real-time optimizations such as in the imbalance settlement compared to model predictive control approaches~\cite{lago2020optimal}, thus further motivating our choice over other alternatives.

Each agent gets trained offline and with a specific battery size in the cluster. By applying different power-to-energy capacity ratios for each virtual battery, we obtain a more differentiated response, making the model more interesting to work with. The MDP used to train the agents is designed to maximize the profit obtained by the batteries.\footnote{Note that this MDP is merely used to train the battery agents for the implicit \emph{response} model. This must not be confused with the MDP defined in \cref{sec:MDP_def}, containing the primary structure used to solve the \emph{price forecast} problem.} The state comprises the battery state (State of Charge and consumed cycles), temporal information, and the last price published by the TSO. 
The action consists of the power withdrawn (injected) by the battery from (to) the grid, and the reward amounts to the profit obtained through the imbalance settlement. To train the agents, we considered historical data publicly granted by the Belgian TSO. In inference, the published prices would be taken from our algorithm, and the actual expenses would then be obtained at the end of each quarter hour based on the price obtained using the formula in \cref{eq:price_formula}. 

%----------------------------------
\subsection{Model of the System}
\label{sec:model_of_the_system}
%----------------------------------
To apply the MCTS technique for publishing imbalance prices, we need a simulable model of the system, \ie a \textit{model that approximates the dynamics of the NRV (SI) over time}. To obtain this, we assume that the NRV dynamics can be partitioned into two distinct and independent parts:
\begin{enumerate}
    \item 
    \label{it:stochastic_variations}
    \textbf{Stochastic variations of the NRV}, caused by exogenous and uncontrollable factors such as weather forecasting inaccuracies. In principle, this part covers all the NRV deviations that are \emph{not} related to implicit (voluntary) reactions of BRPs to the published imbalance prices. 
    \item 
    \label{it:implicit_deviations}
    \textbf{Implicit deviations of the NRV}, caused by the voluntary change in the schedule of BRPs aimed at taking advantage of the imbalance settlement through the published prices.
\end{enumerate}
Following this assumption, we then modeled each part to create a simulable approximation of the system. 
The stochastic variations (\cref{it:stochastic_variations}) were modeled through the usage of a C-VSN forecaster, as explained in~\cref{sec:forecaster}. 
The implicit deviations were instead modeled using a cluster of model-free RL agents, as explained in~\cref{sec:implicit_response}.
The \textit{model of the system} is then obtained by summing the outputs of the two sub-models.
Specifically: given the current time-step $t$, the most recent measurements available are used by the C-VSN forecaster to obtain the expected NRV variations for the next 15 minutes. 
Then, after fixing the price value $\pricePublished_t$ that the TSO will publish in the next timestep, a certain implicit reaction will be triggered from the BRPs. 
This will be modeled by the power value applied by the cluster of RL agents. 
This power value will be summed with the expected NRV obtained from the C-VSN forecaster in the corresponding minute, obtaining the final estimation of the NRV in the next minute. 
The process can be recursively repeated for the next timesteps $\tau \in [t, \tQ]$, each time choosing a different published price $\pricePublished_\tau$ triggering distinct reactions in each step. 
This mechanism will be used by the MCTS algorithm to assess the value of different price publication techniques in the remaining timesteps of the current quarter hour.
% The future NRVs were then predicted by summing the outputs of the two models, namely the expected NRV stochastic deviations given by the forecaster, and the expected NRV deviations from the implicit participation of BRPs (given a certain trajectory of approximated prices published by the TSO, which activates the response of the model).

%-----------------------------
\subsection{MDP definition}
\label{sec:MDP_def}
%-----------------------------
To publish accurate imbalance prices using MCTS, we define the problem as an MDP.
Differently from the MDP defined for the imbalance implicit response, we now define the sequential decision problem of \emph{publishing approximated imbalance prices} from the TSO point of view. 
The MCTS techniques require the MDP to be simulable, \ie the dynamic function of the MDP must be executable by a machine. In the case of a real-world scenario, this is not satisfied, as it is impossible for the TSO to perfectly model and predict the NRV dynamics of the electrical grid. For this reason, we distinguish between two different MDPs:
\begin{itemize}
    \item The \emph{real} MDP:
    $$\left( s, \; a, \; f(s, a, s'), \; \rho(s, a, s') \right ) \in \mathcal{S} \times \mathcal{A} \times [0, 1] \times \mathbb{R} , $$
    that describes the actual (but not fully known) dynamics of the system, and
    \item The \emph{simulated} MDP:
    $$\left( s, \; a, \; \tilde{f}(s, a), \; \rho(s, a, s') \right ) \in
    \mathcal{S} \times \mathcal{A} \times \mathcal{S} \times \mathbb{R} ,$$
    that is composed of simulable models and imitates the actual dynamics of the system.
\end{itemize}
The only difference between the two formulations lies in the transition function, where $f(s, a, s')$ describes the (stochastic) dynamics of the real system and hence is not mathematically formulable; and $\tilde{f}(s, a)$ is a (deterministic) mathematical approximation of the latter and is then executable by a machine. By solving the optimization problem of the simulated MDP we obtain an approximated solution of the original problem. This approximation implies a performance gap, depending on the accuracy of the simulated model compared to the actual dynamics of the system. 

The \emph{state} $s_t$ of the MDP describes the current condition of the grid and contains temporal information, the averaged NRV of the current quarter hour, and the bid ladder for the current quarter hour.

The \emph{action} $a_t$ describes the price published by the TSO at the respective timestep (\ie at each minute). Formally: $a_t = \pricePublished_t \in \mathcal{A} \,\subseteq \mathbb{R}$.
In our solution, we used an MCTS algorithm that requires a discrete action space. We hence discretized the action space in each timestep using its corresponding NRV and bid ladders. For details, see \cref{app:discretization_action_space}. 

The \emph{reward} function describes the general objective of the optimization problem that we are solving. Using an RL-based technique, we are free to choose the reward function that best suits our objectives (\eg price accuracy, balancing costs, NRV reduction). 
We thus experiment with multiple reward functions, each with a distinct mixture of objectives. 
The first reward function is defined as the Mean Absolute Error (MAE) between all the published prices in the current quarter hour and the expected imbalance price (obtained using \cref{eq:price_formula}).
Formally:
\begin{equation}
\label{eq:reward_function_1}
    \rho_1(s_t, a_t, s_{t+1}) = -\frac{\omega_t}{t+1-\tq} \sum_{\tau=\tq}^{t}{\left| \pricePublished_\tau - \priceFormula_{t+1} \right|} \; ,
\end{equation}
where $\omega_t$ is a weighting factor used to increase the relevance of the reward values close to the end of the quarter hour. The intuition behind using $\omega_t$ lies in the fact that the approximation of the prices obtained with \cref{eq:price_formula} gets more reliable as it gets close to the end of the quarter hour. Hence, we assign a higher reward magnitude to the timesteps as they progress forward in the quarter hour. More details about the calculation of $\omega_t$ are presented in \cref{app:omega}.
The sole objective reflected in the reward formula of \cref{eq:reward_function_1} is to predict (and publish), throughout the whole quarter hour, a sequence of prices as close (using MAE as distance) as possible to the real one obtained at the end of the quarter hour: $\priceFormula_\tQ$.
However, from a TSO point of view, different objectives might be integrated into the optimization problem, such as reducing balancing costs or maximizing social welfare.
In particular, as an alternative, we also consider the following extension to the previous reward function:
\begin{equation}
\label{eq:reward_function_2}
    \rho_2(s_t, a_t, s_{t+1}) = -\frac{\omega_t}{t+1-\tq} \sum_{\tau=\tq}^{t}{\left| \pricePublished_\tau - \priceFormula_{t+1} \right|} - \beta_2 \left\Vert \NRV_t \right\Vert_2^2 \; ,
\end{equation}
where $\beta_2 \in \mathbb{R}^+$ is a hyperparameter used to tune the balance between the two objectives of price accuracy and NRV (and thus SI) reduction. 
A third and final reward function that we consider starts from \cref{eq:reward_function_1} and adds a term related to balancing cost. 
We denote the list of activated balancing bids during timestep $t$ by $B_t$, wherein each element is defined as a tuple $(\lambda_\text{act}, V_\text{act}) \in B_t$.
Here, $\lambda_\text{act}$ represents the activated price and $V_\text{act}$ the activated volume.\footnote{For simplicity, we did not consider the ramping factor of the activated power in the calculation.}
We then approximate the balancing cost required in timestep $t$ as follows:
\begin{equation}
    \text{BC}_t = \sum_{(\lambda_\text{act}, V_\text{act}) \in B_t}{\frac{\lambda_\text{act} V_\text{act}}{60}}
\end{equation}
The third candidate reward function thus becomes:
\begin{equation}
\label{eq:reward_function_3}
    \rho_3(s_t, a_t, s_{t+1}) = -\frac{\omega_t}{t+1-\tq} \sum_{\tau=\tq}^{t}{\left| \pricePublished_\tau - \priceFormula_{t+1} \right|} - \beta_3 \text{BC}_{t+1} \; ,
\end{equation}
where $\beta_3 \in \mathbb{R}^+$.
Conceptually similar to the function considered in \cref{eq:reward_function_2}, this third reward formulation is used to optimize the joint problem of price publication accuracy and balancing costs. 

For the sake of a more understandable notation, we presented the reward functions as sums of different physical values (\eg sum of prices [\euro/MWh]
and NRVs [MW]). 
Since these values have different magnitudes we will normalize them into a [0,\,1] range, based on min/max values observed in historical data, before summing them.

\begin{figure}
    \centering
    \includegraphics[width=0.5\textwidth]{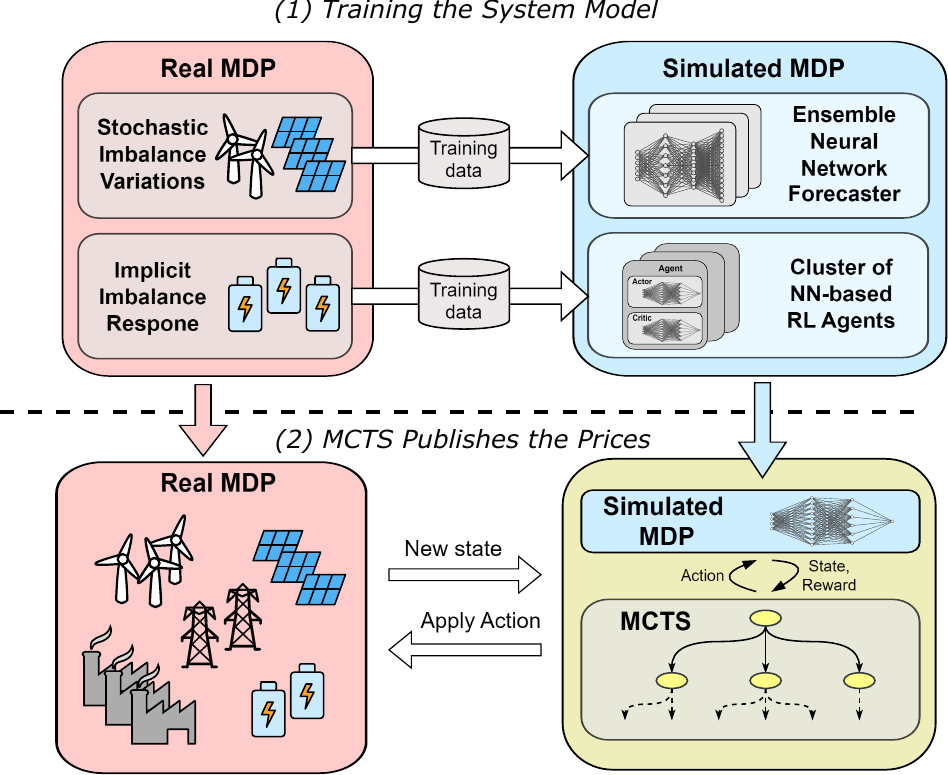}
    \caption{Scheme of our price forecast structure. First (1), a system model is trained using data-driven NN-based techniques. Then (2), the model is used as a simulator paired with an MCTS algorithm to obtain the optimal price value to publish. After publishing the selected action in the real system, the newly obtained state is used to update the simulated environment.}
    \label{fig:action_cycle}
\end{figure}

Last, we describe the \emph{transition} functions of our MDP model. As mentioned earlier, it is unfeasible to perfectly model the actual dynamics of the system. Hence, we approximate them using simulable data-driven tools, as explained in \cref{sec:model_of_the_system}. 
% Specifically, we model the transition function by dividing it into two independent parts, accounting for:
% \begin{enumerate*}[(i)]
%     \item exogenous factors, and\label{it:exo}
%     \item price-dependent reactions of the BRPs.\label{it:brp-reaction}
% \end{enumerate*}
% The first part~\ref{it:exo} regards the stochastic fluctuations of the NRV due to, for instance, weather forecasting inaccuracies. This part is conceptually independent of the published imbalance prices and concerns all the dynamics that are not correlated to the implicit BRPs' reaction in the imbalance settlement. In our work, this part is simulated using the forecaster model described in~\cref{sec:forecaster}.
% The second part~\ref{it:brp-reaction} models the implicit reaction of BRPs to the published imbalance prices, which as described in~\cref{sec:implicit_response} is simulated as a cluster of virtual batteries. 
By using the simulated MDP, we can deploy an MCTS algorithm to obtain a reasonable solution for the environment. The solution (\ie the selected action) will be then executed in the real MDP (\ie the real-world system), which will then generate a new state. The state will be fed into the simulated MDP, which will update its knowledge. The process is then sequentially repeated for each timestep. This cycle is graphically represented in \cref{fig:action_cycle}.

%-------------------------------------
\subsection{Monte Carlo Tree Search}
\label{sec:MCTS}
%-------------------------------------
As introduced earlier, to tackle the imbalance price prediction problem, we deployed an MCTS algorithm, taking inspiration from works such as \cite{silver2018general} and \cite{schrittwieser2020mastering}.
The past decade, MCTS has been integrated with deep learning models, which significantly improved its performance~\cite{silver2017mastering,silver2018general,schrittwieser2020mastering}. Moreover, the method recently got adapted for more complex problems, such as stochastic MDPs~\cite{antonoglou2021planning} and continuous-actions environments~\cite{kim2020monte,lee2020monte}.
However, for the sake of simplicity, we only implement a simple NN-free version to provide a proof-of-concept of the technique. 
Using more advanced algorithm variants is left as future work.

MCTS models environment interactions as a tree, with nodes representing the states (that describe, in our case, the conditions of the grid), and edges modeling the actions (the published prices). 
In the specific problem we are solving, a tree would then describe a list of published price sequences, along with their corresponding impacts on the grid. 
By exploring the environment, the algorithm can evaluate the most promising branches (\ie sequences of actions) and obtain an optimal action for the initial (\ie current) state. Specifically, MCTS iteratively repeats the following steps:
\begin{enumerate}[(1)] %[(i)]
    \item \textbf{Selection:} Starting from the root, the tree gets traversed using a selection criterion that balances between exploration and exploitation, until a current leaf node is reached.
    \item \textbf{Expansion:} The leaf node gets expanded, adding new nodes to it.
    \item \textbf{Backpropagation:} Starting from the expanded node, the information concerning the selected trajectory's rewards gets backpropagated to the root node, updating each node's knowledge along the way.
\end{enumerate}
The tree gets initialized as a single root node describing the initial system state. Then, these phases are repeated a sufficient amount of time.
Next, we provide the details of each phase of our implementation, largely based on~\cite{schrittwieser2020mastering}.\footnote{The implementation is mainly the same, except for the usage of neural networks which has been completely removed in our implementation, and the removal of the logarithmic portion of the selection equation.}
We use a superscript to indicate the depth of the nodes and actions in the selected branch, starting from 0 (root) up to $\ell \in \mathbb{N}$ (leaf). Moreover, due to the discretization of the action space described in \cref{app:discretization_action_space}, each node $s^k$ will have a distinct action space $\mathcal{A}^k$.
\begin{enumerate}[(1)]
    \item \textbf{Selection:} Starting from the root node $s^0$, the current tree is traversed by selecting an action (edge) $a^k \;, \forall \; k  = 0, 1, \dots, \ell-1$ until a leaf node $s^\ell$ is reached. The selection is based on the following formula:
    \begin{equation}
        a^k = \arg \max_{a \in \mathcal{A}^k} \biggl\{ \overline{Q}(s^k, a) + \alpha \frac{\sqrt{N(s^k)}}{1+N(s^k, a)} \biggl\} \;\; ,
    \end{equation}
    where $N(s) \in \mathbb{N}$ is the number of visits received by node $s$, $N(s, a) \in \mathbb{N}$ is the number of times action $a$ has been selected from state $s$, $\alpha \in \mathbb{R}^+$ is a hyperparameter used to balance exploration and exploitation, and $\overline{Q}(s, a)$ is the normalized version of a state-action value function approximation that is tuned in the backpropagation phase.
    \item \textbf{Expansion:} When a leaf node $s^\ell$ is reached, it gets expanded with new nodes (one for each action in $\mathcal{A}^\ell$), thus increasing the depth along the followed path.
    \item \textbf{Backpropagation:} After the expansion phase, the rewards obtained in the selected branch get backpropagated to the root. Specifically, starting from the newly expanded node $s^\ell$, the state-action value function gets updated:
    \begin{align}
        Q(s^k, a^k) = \frac{N(s^k, a^k) \> Q(s^k, a^k) + G^{k+1}}{N(s^k, a^k) + 1} \; ; \\ 
        \nonumber \forall k = \ell-1, \dots, 0
    \end{align}
    where $Q(s, a)$ is initialized as $\rho(s, a)$ and $G^k$ is the discounted accumulated reward defined as:
    \begin{equation}
    %\label{eq:mcts_accumulated_reward}
        \begin{cases}
          G^\ell = \rho(s^{\ell-1}, a^{\ell-1}) \\
          G^k = \rho(s^{k-1}, a^{k-1}) + \gamma \> G^{k+1} %\; ; \forall
          \quad \forall k = \ell-1, \dots, 1
        \end{cases} \nonumber
    \end{equation}
    where $\gamma \in [0, 1]$ is the discount factor. The normalized version is obtained through 
    min-max scaling based on the values in the current tree:
    \begin{equation}
        \overline{Q}(s, a) = \frac{Q(s, a) - Q^\text{min}}{Q^\text{max} - Q^\text{min}} \in [0, 1] \;\; ,
    \end{equation}
    where 
    \begin{equation*}
        \begin{cases}
            Q^\text{min} \doteq \min_{s', a' \in \text{Tree}}{Q(s', a')} \\
            Q^\text{max} \doteq \max_{s', a' \in \text{Tree}}{Q(s', a')}
        \end{cases}
    \end{equation*}
    Moreover, the number of visits of each node in the trajectory gets updated:
    \begin{equation}
        \begin{cases}
        N(s^k, a^k) = N(s^k, a^k) + 1 & \quad \forall k = \ell, \dots, 0 \\
        N(s^k) = N(s^k) + 1 & \quad \forall k = \ell, \dots, 0 
        \end{cases} \nonumber
    \end{equation}
\end{enumerate}
A single execution of these three phases is referred to as a \emph{simulation}. By increasing the number of simulations involved, a more optimal solution (\ie an action that further maximizes future rewards) is obtained. After a sufficient amount of simulations has been performed, the action $a^*$ is selected as the one that has the higher state-action score: ${a^* = \argmax_{a \in \mathcal{A}^0}{Q(s^0, a)}}$.

%==============================================
\section{Experiment Setup}
\label{sec:experiment-setup}
%==============================================

\subsection{Data Used and Validation Metrics}
\label{sec:data_and_metrics}
To train our system dynamics model, we used historical data of the Belgian grid.\footnote{Publicly available in Elia's website~\cite{eliaopendata}.} The SI forecaster described in \cref{sec:forecaster} is trained using data from 2020 to 2022. The implicit response model (\ie comprising the batteries' controllers) is trained and validated using data from 2022. To simulate the implicit response of BRPs, we used a cluster of four batteries, each with a different capacity ratio. 
Moreover, to prove the method efficacy with varying magnitudes of response, we considered different battery sizes: a \emph{small} one with a total capacity (sum of every battery capacity) of 60\,MW\,--\,150\,MWh, a \emph{medium} one with a total capacity of 125\,MW\,--\,310\,MWh, and a \emph{big} one with a total capacity of 250\,MW\,--\,620\,MWh. 
Please note that, although the battery sizes were chosen based on plausible capacities,\footnote{For example, we can consider current projects such as https://www.energy-storage.news/continental-europes-biggest-battery-system-in-inaugurated-by-corsica-sole-in-belgium/ and https://www.nyrstar.com/resource-center/news/virtual-battery} their exact specifications are not particularly relevant to our experiments. Our research aims to demonstrate that our method produces valuable results, even when confronted with varying response magnitudes. Hence, we limited our experiments to select three significantly different magnitudes of response. We specifically chose power capacities to encompass various multiples of the average absolute NRV observed in historical data. A more detailed study on the matter, although beyond the scope of this paper, should be conducted in the future. 
Moreover, to extend the experiments towards more significant (and currently unrealistic) response magnitudes, we also considered different cases with a single-battery response of 500\,MW\,--\,1,000\,MWh, up to 2,000\,MW\,--\,4,000\,MWh (\cref{sec:results_exp1.5}).

\begin{figure*}
    \centering
    \includegraphics[width=1\textwidth]{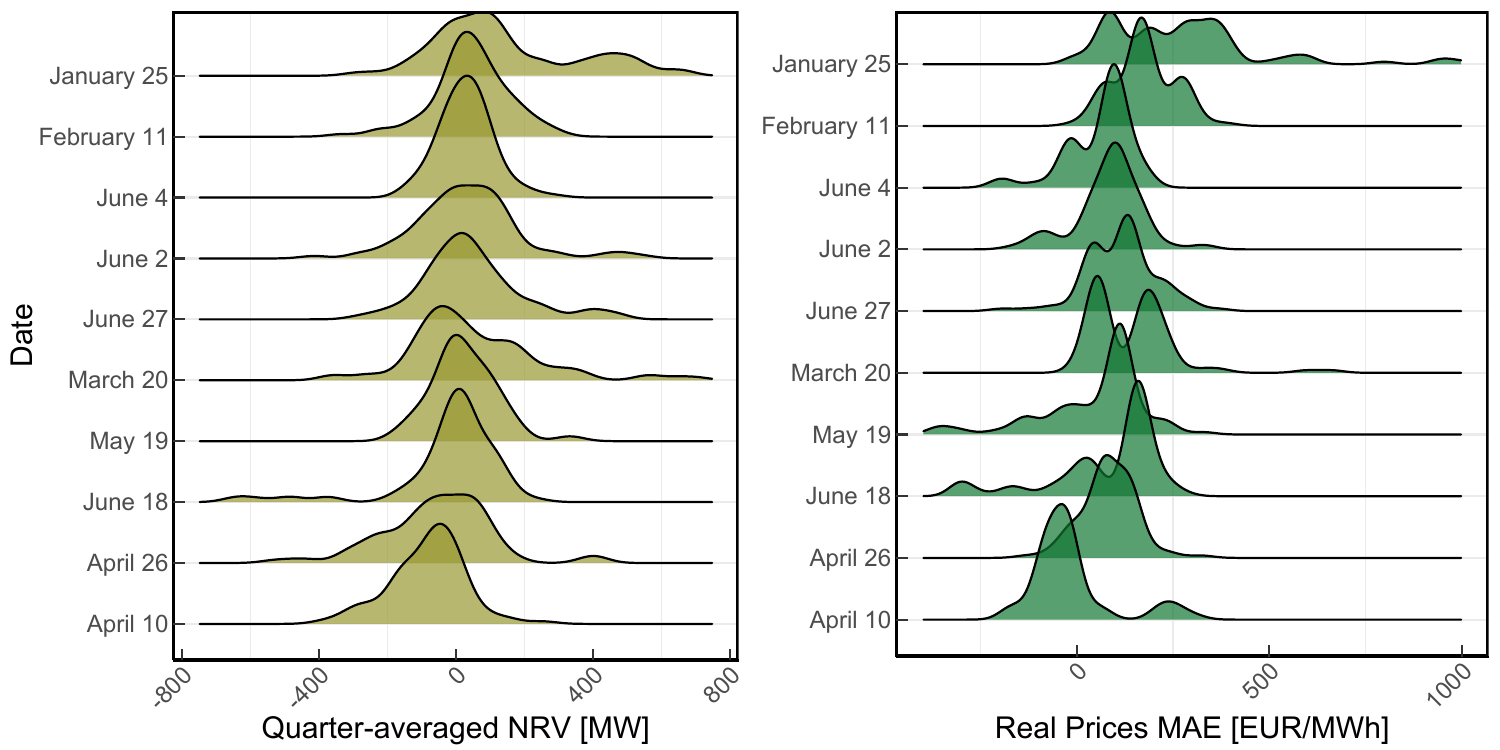}
    \caption{Distribution of the quarter-averaged NRV (left) and of the applied imbalance prices (right) for the evaluation days chosen for our experiments. The days were sampled from different clusters obtained using a $k$-means clustering technique based on each day's NRV and price values. Please notice that, for graphical reasons, the x-axis range has been trimmed and does not necessarily reflect the full price range. For example, even if for only a few quarter hours, on January 25, we observed prices of over 3,000~\euro/MWh.}
    \label{fig:nrv_prices_distribution}
\end{figure*}

The experiments are evaluated for 10 days that we sampled from the period January 2023 to June 2023. More specifically, to select the days used for evaluation, we used the $k$-means clustering algorithm to partition the full dataset into 10 groups based on the average imbalance price, the average NRV, the standard deviation of the NRV, and the standard deviation of the (historically) published prices.
Every evaluation day is then sampled from different clusters, ensuring sufficient diversity in the evaluation set.\footnote{To preserve a fair comparison between each experiment, we fix the evaluation set of the same 10~days across all runs.} The NRV and price distributions of the sampled days can be observed in \cref{fig:nrv_prices_distribution}. 
The sampled days are considered using the Coordinated Universal Time (UTC) timezone. 

To keep the running time of our experiments acceptable, we limited the tree search to a maximum of 200 simulations. After different trials, we identify such a hyperparameter to be a compromise between computational cost (with a running time of about 10\,s for MCTS to determine the eventual action, \ie price publication, for each timestep) and algorithmic performance. However, in realistic situations, the TSO could further increase the number of simulations to fully exploit the time available in each minute-based prediction. Therefore, the results presented here are to be taken as a lower bound regarding results effectiveness, as more simulations might improve the algorithm results.

To evaluate our experiments, several metrics and objectives have to be taken into consideration to evaluate the MCTS technique. The first metric used is the Mean Absolute Error (MAE) between the prices published within the quarter hour and the actual price obtained at the closure of the settlement period. This metric is used to indicate the price accuracy of the method. 
To analyze the effect of the published price on the grid status, we also kept track of some grid-related values such as the minute-based NRV magnitudes and the costs required by the TSO for balancing purposes. Moreover, we also analyzed additional values that might be relevant for both the TSO and the BRPs, such as the revenue obtained by imbalance settlement participants, and the variability of the prices published within the same quarter hour. These metrics are shown in \cref{app:full_metric_list}.

\subsection{Experiments Formulation}
To perform our analyses, we ran four different groups of experiments, each with different objectives:
\begin{E}
    \item 
    \label{exp:experiment1}
    To demonstrate the potential of our approach as a proof-of-concept, we assess its performance under ideal conditions, \ie considering perfect knowledge of both the NRV fluctuations and the implicit imbalance response model. In these experiments, we only optimize the price accuracy (using \cref{eq:reward_function_1}) and we benchmark the results over the current publication method while considering different reasonable battery sizes (\cref{sec:results_exp1}). 
    \item 
    \label{exp:experiment1.5}
    Next, we study the price publication performance in case of higher response magnitudes, by considering more extreme cases up to a 2GW/4GWh battery response capacity (\cref{sec:results_exp1.5}). 
    \item 
    \label{exp:experiment2}
    Subsequently, we study the ability of our technique to optimize multiple objectives (\ie increasing the price accuracy, decreasing the SI magnitude, and decreasing the balancing costs)(\cref{sec:results_exp2}). To this end, we extend the experiments performed in \textbf{E1} with different reward functions (\cref{eq:reward_function_2,eq:reward_function_3}), using a medium-sized battery as response model. 
    \item
    \label{exp:experiment3}
    Finally, we analyze the performance of our technique in more realistic settings, thus gauging its practical applicability for TSOs in the real world. Instead of assuming perfect knowledge of the NRV fluctuations, we implemented a forecaster as described in \cref{sec:forecaster}. Moreover, we added artificial noise to the implicit response model; evaluating the technique using different magnitudes of model inaccuracies (\cref{sec:results_exp3}).
\end{E}

%==========================
\section{Results}
\label{sec:results}
%==========================
\subsection{Price Accuracy under Ideal Conditions}
\label{sec:results_exp1}
We first show the results for the experiments described in \textbf{E1}, using ideal conditions (perfect NRV forecast and response knowledge). Specifically, we consider four different experiments, each using different magnitude sizes of the response battery. We then compared the price publication accuracy of our technique with the one obtained using the considered baseline. The results are shown in \cref{tab:results_exp1}.

\begin{table*}[t]
\centering
\captionof{table}{Results obtained when considering perfect conditions. The MCTS-based technique achieves more accurate predictions of the imbalance prices, with better results for higher implicit response magnitudes.
Moreover, the technique does not negatively affect the NRV and the Balancing costs.}
    \begin{tabular}{l|rrr|rr|rr}
        
        \multicolumn{1}{c}{} &
        \multicolumn{3}{c}{\textbf{Published Prices MAE [€/MWh]}} & 
        \multicolumn{2}{c}{\textbf{NRV MAE [MW]}} &
        \multicolumn{2}{c}{\textbf{Balancing costs [€]}}
        \\
        \toprule
        \textbf{Response Magnitude} &
        \textbf{MCTS} & \textbf{Baseline} & \textbf{Improvement [\%]}
        & \textbf{MCTS} & \textbf{Baseline} & \textbf{MCTS} & \textbf{Baseline}
        \\
        \midrule
        No response 
        & 18.68 & 22.47 & 16.87\,\% & 129.81 & 129.81 & 3,439.06 & 3,439.06
        \\
        Small battery 
        & 18.83 & 23.54 & 20.01\,\% & 123.68 & 124.77 & 3,353.07 & 3,375.07
        \\
        Medium battery 
        & 18.62 & 23.14 & 19.54\,\% & 119.51 & 120.82 & 3,292.99 & 3,327.06
        \\
        Big battery 
        & 18.05 & 22.69 & 20.42\,\% & 119.02 & 120.03 & 3,332.02 & 3,347.83
        \\
        \bottomrule
    \end{tabular}
\label{tab:results_exp1}
\end{table*}

\begin{figure*}
    \centering
    \includegraphics[width=1\textwidth]{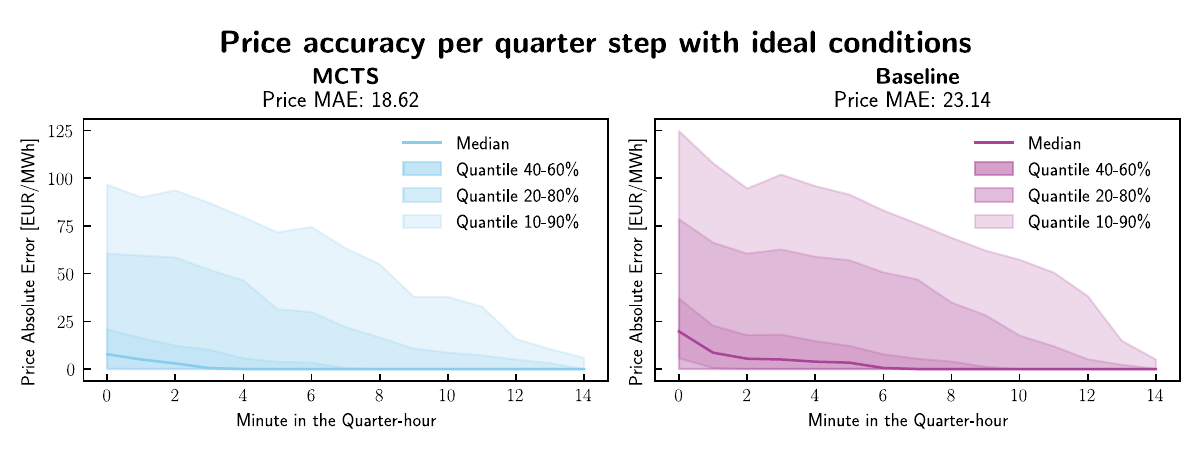}
    \caption{Price publication error quantiles for each minute of the quarter hours considered in the experiment (960 in total from the 10 days sampled as explained in \cref{sec:data_and_metrics}) with a medium response magnitude and ideal conditions. The MCTS technique obtains substantially lower error values throughout the whole quarter hour. 
    The improvements are more noticeable in the first minutes of the quarter hours. We speculate that this is due to the cumulative average mechanics of the NRV used to obtain the price publication for the baseline technique.}
    \label{fig:price_quantiles}
\end{figure*}

We observe a substantial improvement in the published price accuracy of the MCTS-based method, up to $+$20.4\%. This result is not surprising, as perfect knowledge of the model enables the tree search to accurately identify the best actions in relation to the market responses and the expected NRV fluctuations. In contrast, the baseline obtains the published prices by only looking at the current NRV, thus only using incomplete information.
In addition to prediction quality, we also tracked the averaged NRV and Balancing costs (as described in \cref{sec:MDP_def}).
We observe that the MCTS-based technique does not negatively affect these values, which remain almost invariant compared to the ones obtained using the Baseline.
A more detailed analysis of MCTS's ability to reduce the NRV and the balancing costs is provided in \cref{sec:results_exp2}, while \cref{app:full_metric_list} reports on additional metrics of interest.
We now show a more in-depth analysis of the prediction technique.
\Cref{fig:price_quantiles} shows different quantiles for the absolute error of the published prices in every minute of the quarter hour, to analyze how prediction accuracy varies within a quarter hour. 
These results assume a medium response magnitude.

\Cref{fig:price_quantiles} shows that the MCTS error quantiles are, overall, lower than the Baseline ones for all minutes of the quarter hour. This is in line with the results shown in \cref{tab:results_exp1}. 
The first few minutes of each quarter hour exhibit the most noticeable improvements, which aligns with how the NRV is used for predictions. The NRV in the imbalance formula is based on minute-level measurements cumulatively averaged over the current quarter hour (as shown in \cref{eq:average_SI}). This means that early NRV values have a significant impact on the cumulative average, as they are averaged over fewer data points. In contrast, as the quarter hour progresses, additional NRV measurements have less effect on the cumulative average. Since the baseline method relies solely on this cumulative average, its predictions are less accurate at the start of the quarter hour, when the average is still highly sensitive to new measurements. The MCTS-based method, however, improves accuracy by considering both the current average NRV and future NRV projections, which tend to be more stable and reliable if the predictions are accurate. As a result, this technique performs notably better than the baseline, especially in the early minutes of the quarter hour when the available information still lacks robustness and reliability.

\begin{figure*}
    \centering
    \includegraphics[width=1\textwidth]{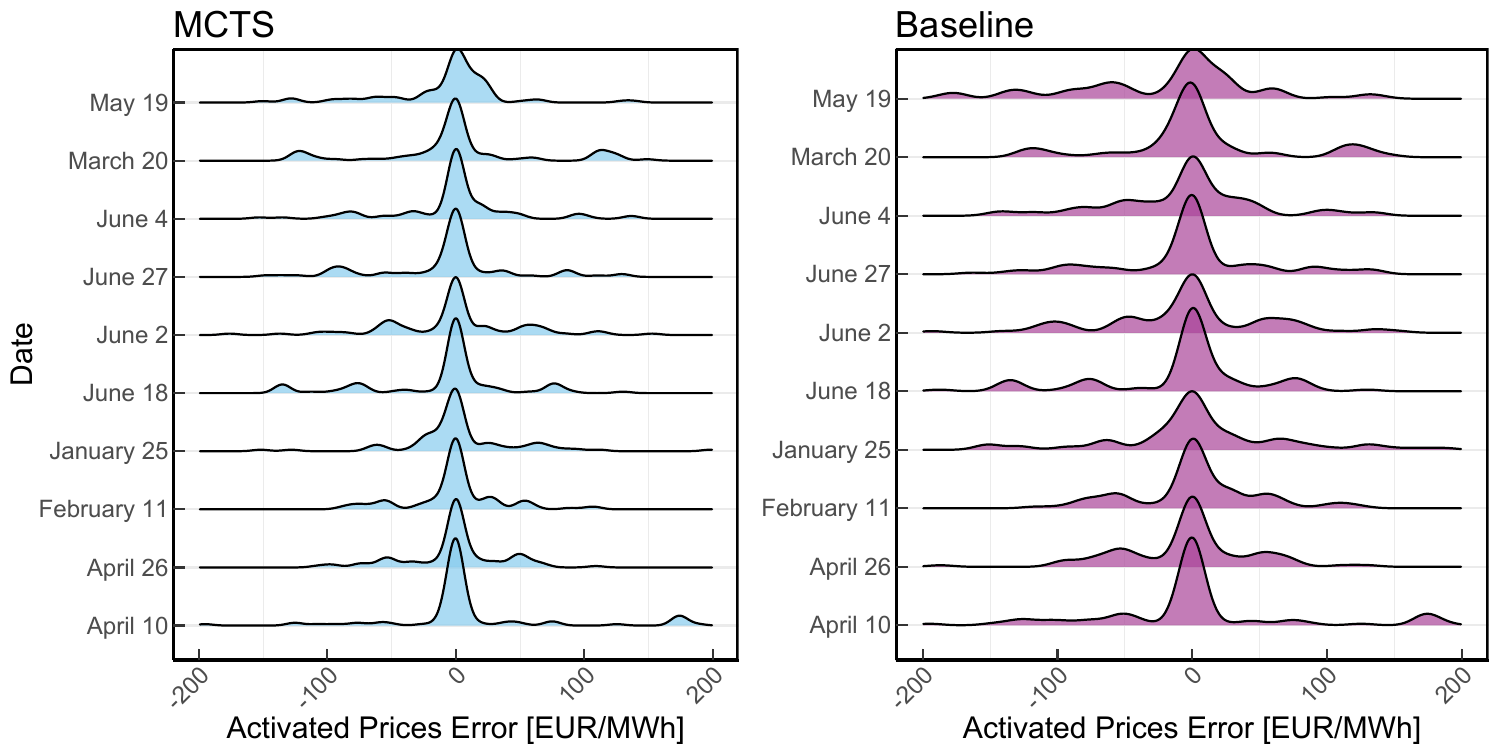}
    \caption{Distribution of the error between the published prices and the applied one for the MCTS-based technique vs. the baseline purely using \cref{eq:price_formula}. A positive error means an overestimation of the published prices compared to the final one. The MCTS-based technique achieves a more narrow and centered distribution mass around an error value of 0 (\ie perfect predictions). Notice that, for a more informative graph, the plotted x-axis values are limited to the interval of $[-200, 200]$.}
    \label{fig:prices_distribution_comparison}
\end{figure*}

To further investigate the techniques' performances, \cref{fig:prices_distribution_comparison}  shows the price error distributions for each of the 10 sampled test days (as explained in \cref{sec:data_and_metrics}).
The MCTS-based technique has a more narrow distribution of error around the value of 0 (\ie perfect predictions). Moreover, the graph shows that the technique does not provide sporadic publications that are particularly inaccurate, but its distributions are rather an improved version of the ones obtained through the baseline. 

\begin{figure*}
    \centering
    \includegraphics[width=1\textwidth]{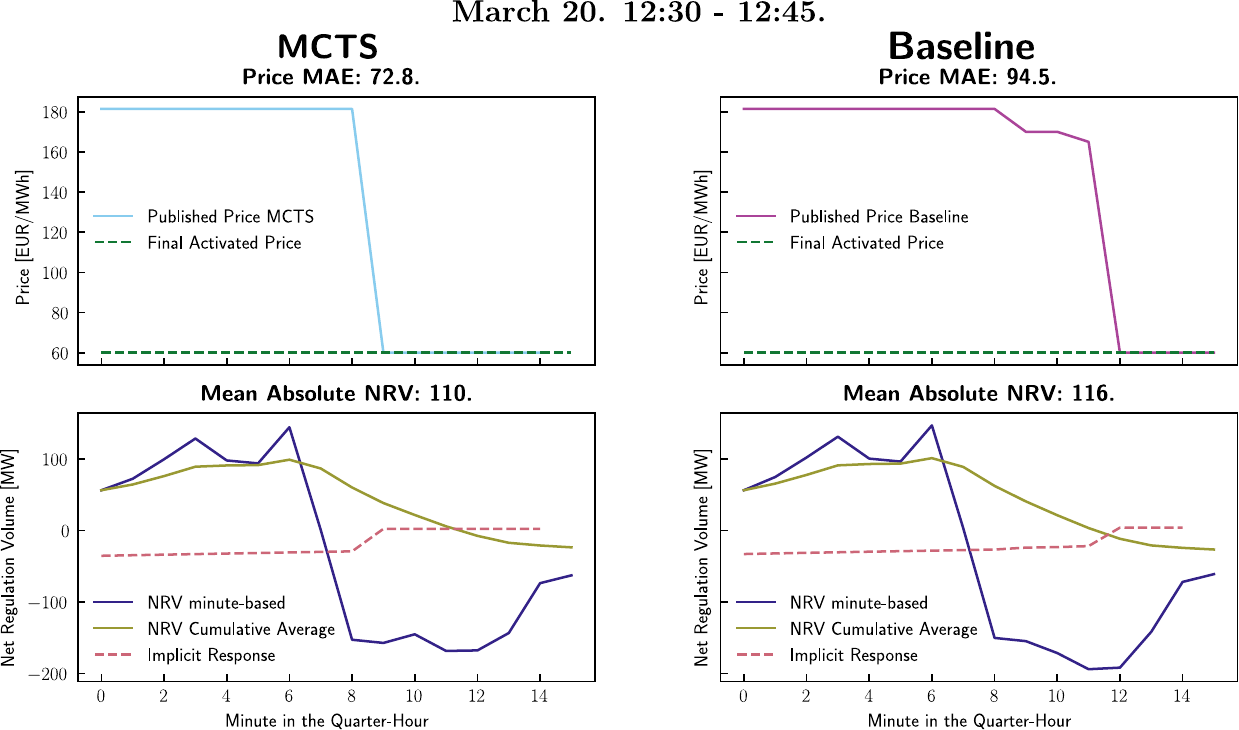}
    \caption{Minute-based plot of the publications obtained in a quarter hour for MCTS (left) and the Baseline price formula (right). The implicit response activations shown in the second row graphs are indicated from the BRP point of view (\ie positive activations mean that the batteries are charging, increasing the NRV, and vice versa for negative activations). Also, please note that the implicit activations influence the minute-based NRV the minute after they are shown.}
    \label{fig:daily_plot}
\end{figure*}

Last, we provide a graphical view of the techniques' predictions. \Cref{fig:daily_plot} shows an example of the MCTS algorithm publications vs. the one obtained using \cref{eq:price_formula} for a medium-sized battery response in a selected quarter hour.
In this quarter, the measured NRV fluctuates significantly in both sign and magnitude. Initially, it is positive, but around minute 8, it abruptly shifts to negative and remains so until the end of the period. As a result, the cumulative average NRV -- which strongly influences the final imbalance price -- starts with a high positive value but gradually decreases, ultimately settling near zero on the negative side.
The instability of the sign particularly complicates the problem of publishing accurate prices in this specific quarter. 
Ideally, the publication technique should forecast, at each timestep, the final average NRV for the quarter hour. Based on this prediction, it should then publish prices that align with the expected final price. Moreover, as an additional challenge, the published prices themselves influence the NRV, triggering implicit responses that push it in the opposite direction. While these responses often reduce the NRV's magnitude, they can also overshoot, amplifying fluctuations and even reversing the NRV sign. Therefore, the technique must not only predict accurately but also strategically adjust published prices to prevent these responses from distorting the final average NRV.

In \cref{fig:daily_plot}, our proposed MCTS algorithm effectively anticipates price changes by predicting and publishing expected future prices before the cumulative average NRV reaches the price-change threshold. Compared to the baseline approach, which only updates prices reactively when the threshold is crossed, MCTS accounts for upcoming NRV shifts and adjusts accordingly. This proactive approach improves accuracy and stability in price publications. 
% Furthermore, increasing the number of MCTS simulations could enhance its predictive capability, allowing deeper exploration of possible NRV fluctuations and further refining price adjustments.

% The proposed MCTS algorithm is able to look forward for the changes of magnitude in the NRV, effectively anticipating price changes and obtaining more accurate publications. 
% In \cref{fig:daily_plot}, our proposed MCTS algorithm effectively anticipates price changes by predicting and publishing expected future prices \emph{before} the cumulative average NRV reaches the price-change threshold.
% Instead, simply using \cref{eq:price_formula} does not handle such scenarios (the published price only changes when the NRV surpasses its price-changing threshold, \ie becomes negative in \cref{fig:daily_plot}).
Even though in the example the MCTS algorithm anticipates the price change only a few timesteps in advance, the resulting improvement in price accuracy remains significant. 
It is in fact unrealistic to expect the technique to predict such shifts from the outset of the quarter, particularly in complex scenarios such as the one presented. However, increasing the number of MCTS simulations could potentially further enhance performance by enabling deeper tree searches, effectively extending the algorithm’s predictive horizon and improving its ability to anticipate price fluctuations.

\subsection{Publication Performance When Dealing With High Response Magnitudes}
\label{sec:results_exp1.5}

\begin{figure}
    \centering
    \includegraphics[width=0.5\textwidth]{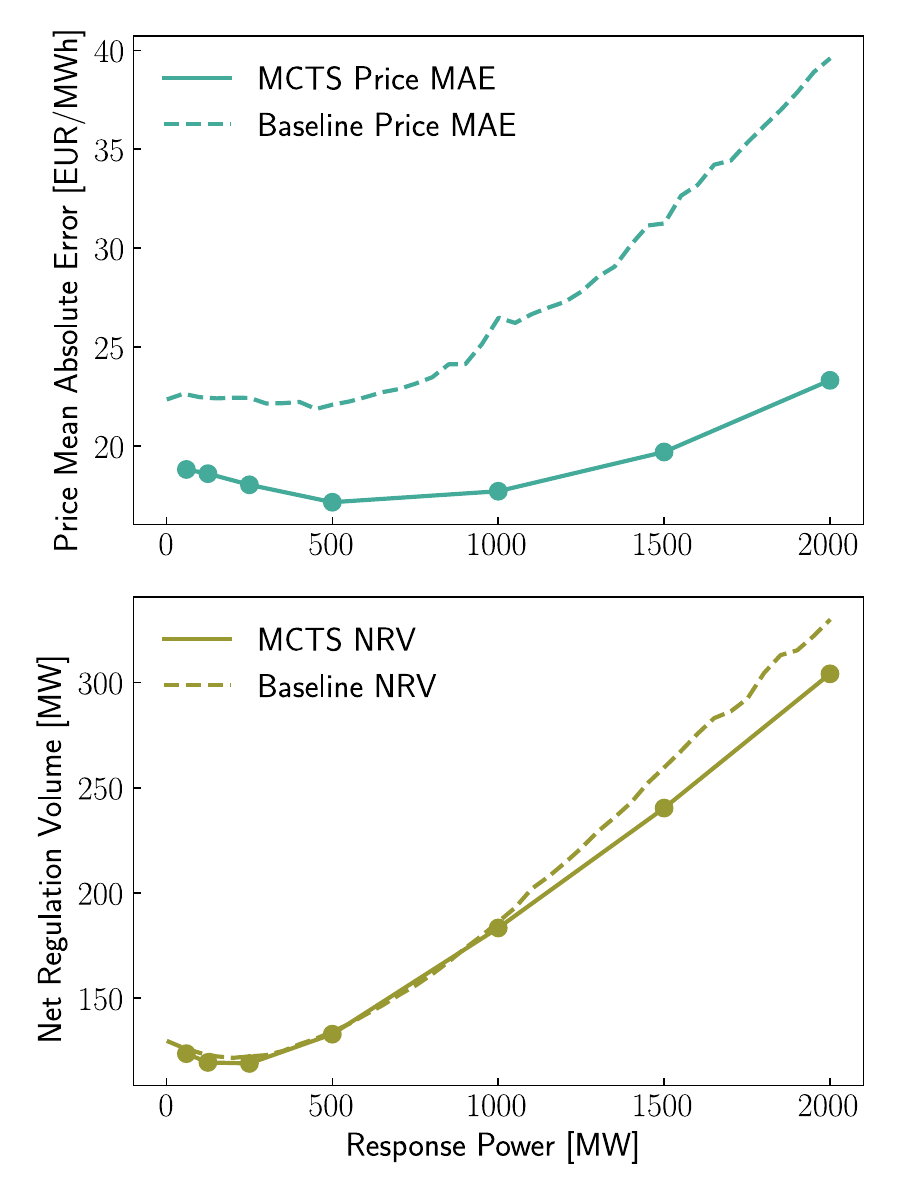}
    \caption{Results when dealing with an increasing response magnitude in the grid, up to 2GW of response power. The MCTS-based publication technique performs better for both the average NRV measured and the published price accuracy.}
    \label{fig:high_responses_graph}
\end{figure}

To analyze the plausible consequences of a significantly higher response capacity in the grid, we extended the experiments of \cref{sec:results_exp1} by increasing the power capacity of the responses up to 2\,GW. 
\Cref{fig:high_responses_graph} shows the 
resulting publication accuracy and average absolute NRV for both the baseline and the MCTS-based techniques.
Once again, the reward function considered is \cref{eq:reward_function_1}.
We note that the NRV magnitude measured in the grid increases when it surpasses a certain threshold (around 250\,MW). 
This is somehow expected, as a high response power would provide an excessive injection to (or an excessive withdrawal from) the grid. For example, assuming a positive NRV at a certain moment, following \cref{eq:price_formula} a relatively high price would be published, causing the BRPs to sell energy to the grid as a reaction. 
However, in case of an excessive response, the injected power would change the NRV sign, therefore contributing to grid instability. 
A similar reasoning can be applied to the price accuracy, as such excessive reactions would cause the NRV sign to be unstable. As a consequence, the baseline technique would have high oscillations in their publications, with a consequential increment in the publication errors. On the other hand, the proposed MCTS-based technique should be able to predict such behaviors and thus try to avoid these instabilities with more strategic publications. We indeed observe this in the graph in \cref{fig:high_responses_graph}, where publication errors are significantly lower when using the MCTS-based technique, compared to the baseline. 
MCTS also attains a slightly lower average NRV, although not significantly. That might be due to the fact that the NRV is not being directly optimized, since it is not represented in the reward function. Hence, we next explore the impact of different reward functions.

\subsection{Reward Functions for Multi-objective Optimization}
\label{sec:results_exp2}
We now analyze more advanced reward functions (namely \cref{eq:reward_function_2,eq:reward_function_3}), which aim to jointly optimize multiple objectives.

\subsubsection*{NRV reduction + price prediction}
\begin{figure}
    \centering
    \includegraphics[width=0.5\textwidth]{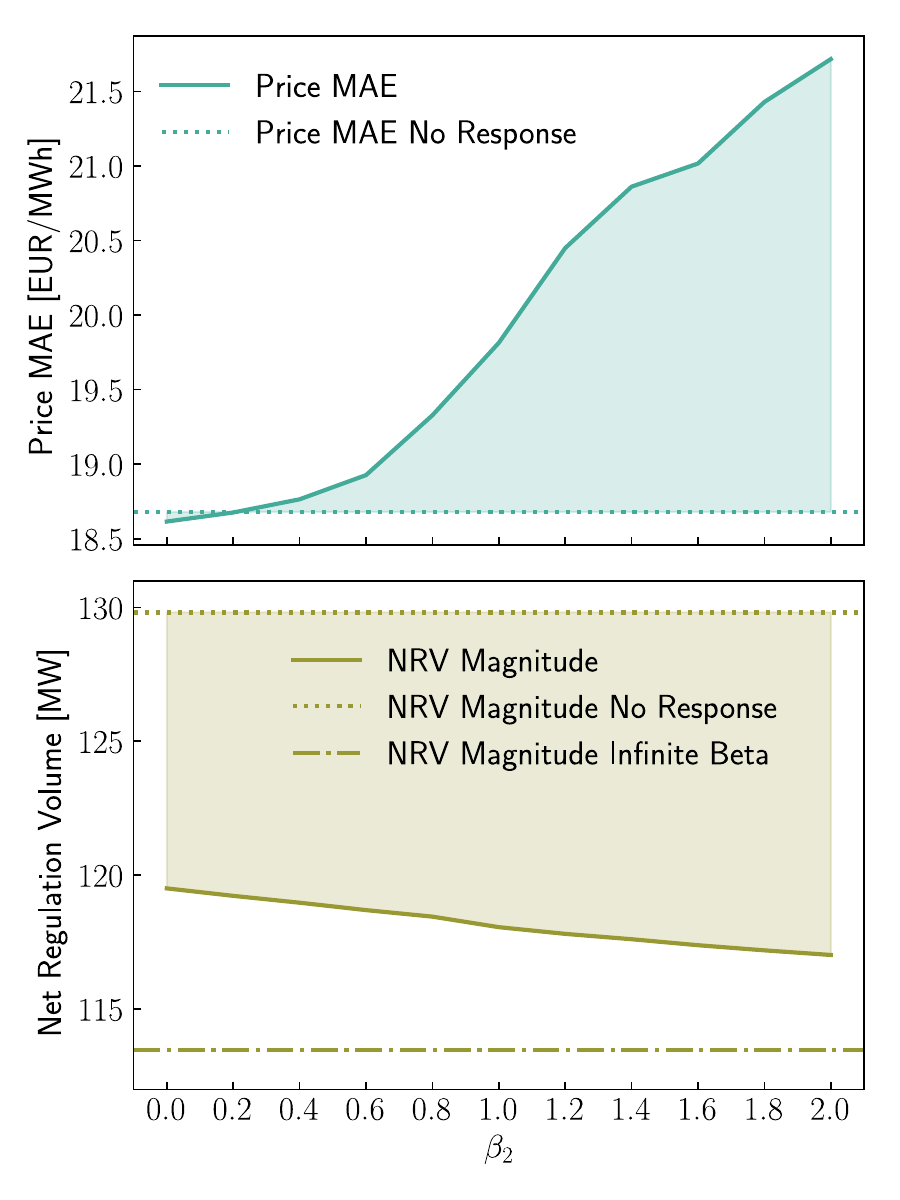}
    \caption{Analysis of the possible NRV reduction achievable by the considered battery (with a size 
    in the range of 125\,MW\,--\,310\,MWh) vs. the loss in publication accuracy of the prices. The graph compares the obtained NRV reduction with different values of $\beta_2$ against the case where $\beta_2 \rightarrow +\infty$ (most NRV reduction scenario) and the case where no response is involved (no control over the NRV). Despite obtaining a reduction in averaged NRV magnitude by increasing $\beta_2$, most of the reduction is achieved by just having a response that follows accurate prices.}
    \label{fig:beta1_analysis}
\end{figure}

\noindent We first focus on adding \emph{NRV reduction} to the price prediction objective, \ie using \cref{eq:reward_function_2}.
\Cref{fig:beta1_analysis} plots the published prices MAE and the absolute average NRV when varying the $\beta_2$ parameter (higher values of $\beta_2$ increase the relevance of the NRV reduction factor in the reward function). 
% \begin{figure}
%     \centering
%     \includegraphics[width=0.48\textwidth]{price_acc_vs_nrv_reduction}
%     \caption{Results obtained when varying $\beta_2$ in the reward function defined in \cref{eq:reward_function_2}.}
%     \label{fig:results_reward_funct_2}
% \end{figure}
From the graph we observe that, as expected, the technique is able to reduce the NRV magnitude when higher values of $\beta_2$ are used, sacrificing accuracy in the price publications.
While this is what we expected the algorithm to achieve and therefore enables TSOs to design their objectives more broadly, the NRV reduction obtained may not necessarily be cost-effective.
Indeed, the relative NRV reduction ($\sim$\,2.8\% for $\beta_2=2$) is rather low compared to the relative loss of accuracy in price prediction ($\sim$\,14.0\% for $\beta_2=2$).
This was to be expected, as the ability of the flexible BRP assets to reduce the NRV magnitude is, when averaged daily, relatively limited.
This limitation is mostly due to the cycle constraint enforced on the responsive batteries, which limits the times per day they can participate in NRV reduction.
To quantify these limits (or, in other words, to assess the average NRV reduction the considered response model can achieve at most), we consider the case where no response is involved (\ie the published prices do not affect the NRV in any way). 
The averaged NRV obtained in this case corresponds to the midpoint of the interval that defines the range of NRV reduction (or increase) achievable by the given battery capacities.

To approximate the maximal NRV reduction the response model can achieve, we run our simulations only using the NRV reduction part of the reward formula in \cref{eq:reward_function_2} (which conceptually corresponds to the converged value when $\beta_2 \rightarrow +\infty$).
% Similarly, for reference purposes, we also assess the minimal NRV reduction (achieved by $\beta_2 \rightarrow -\infty$; other than providing a reference point, this strategy is not particularly meaningful for the scope of this paper).
The converged value is shown in \cref{fig:beta1_analysis}, thus showing the achievable NRV gap for varying $\beta_2$.
From the graph, we observe that most of the possible NRV reduction is already achieved by having a bigger response (or, in this case, a response at all) following the published prices, independently of the NRV reduction term in the reward function. While a bigger reduction can be achieved with a higher $\beta_2$ (up to an additional reduction of $\sim$8\,MW on average), the published price error, in that case, would dramatically increase (from a published price MAE of 18.62\,\euro/MWh with $\beta_2$\,=\,0 to a MAE of 53.19\,\euro/MWh when the reward function only considers the NRV reduction as objective). 

\subsubsection*{Balancing cost reduction + price prediction}
\noindent Our analysis of aiming for joint minimization of balancing costs and price prediction errors, \ie using \cref{eq:reward_function_3}, leads to similar conclusions to the ones drawn for the NRV reduction term in \cref{eq:reward_function_2}.
\begin{figure}
    \centering
    \includegraphics[width=0.5\textwidth]{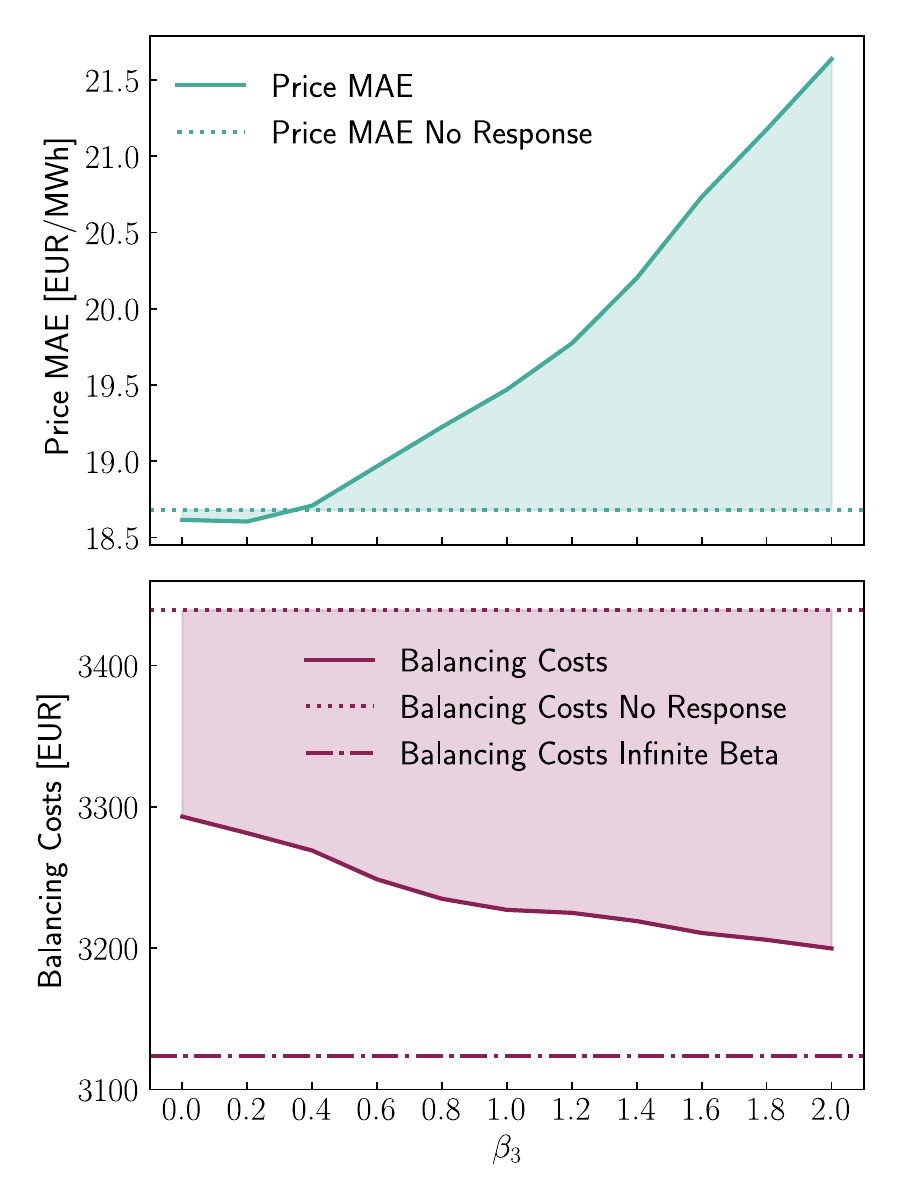}
    \caption{Analysis of the possible balancing costs reduction achievable by the considered battery (with a size of 125MW\,--\,310MWh) vs. the loss in publication accuracy of the prices. The graph compares the obtained balancing cost reduction with different values of $\beta_3$ against the case where $\beta_3 \rightarrow +\infty$ and the case where no response is involved.}
    \label{fig:beta2_analysis}
\end{figure}
As for the NRV+price case, the relative reduction of the newly added metric, in this case balancing cost, is relatively low compared to the rise in price prediction error.
A similar analysis of the $\beta_3$ convergence is shown in \cref{fig:beta2_analysis}. 
Similar to the NRV case, an extra reduction in balancing costs can be obtained by further increasing $\beta_3$, but with a {significant increase in MAE of the predicted prices, \ie up to 64.4\,\euro/MWh when $\beta_3\rightarrow\infty$.
\\
In conclusion, TSOs can indeed achieve a stronger reduction in NRV or balancing costs by adding corresponding penalty terms to the reward function. 
However, the reductions obtained are not necessarily cost-effective.
TSOs should thus carefully assess the obtained benefits with the price accuracy sacrificed to strike an acceptable balance.

\subsection{Results in More Realistic Conditions}
\label{sec:results_exp3}

\begin{figure*}
    \centering
    \includegraphics[width=1\textwidth]{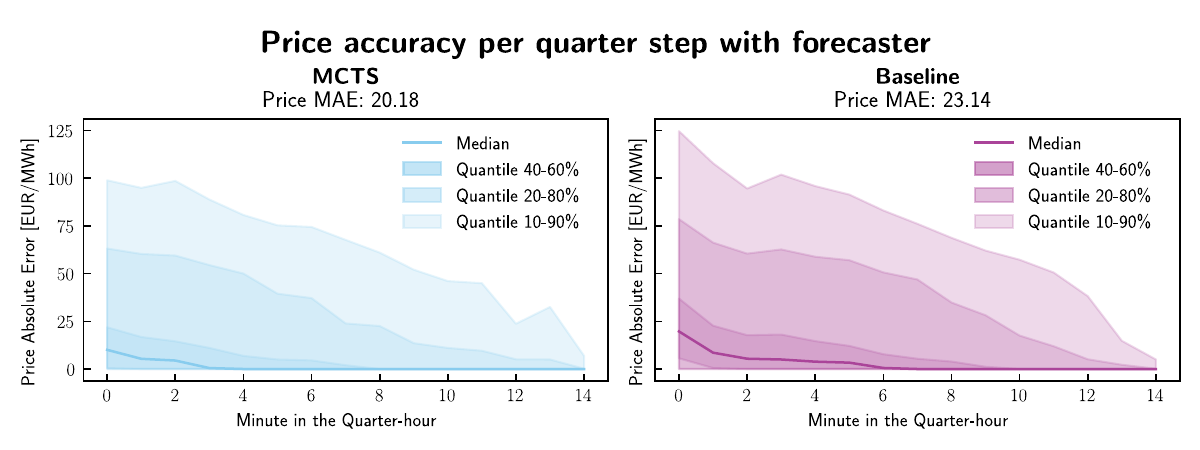}
    \caption{Price publication error quantiles in each minute of the quarter hours considered in the experiment (960 in total) with a medium response magnitude and using the NRV forecaster. The results still assume perfect knowledge of the response model. As in the case with ideal conditions, the MCTS is able to consistently outperform the baseline technique, with a decrement of the published prices MAE of 12.8\%.}
    \label{fig:price_accuracy_quantile_forecaster}
\end{figure*}

Finally, we show the results of our proposed technique when dealing with more realistic conditions. 
First, we analyze the effect of imperfect NRV forecasting, by using the NRV forecaster of \cref{sec:forecaster}.
\Cref{fig:price_accuracy_quantile_forecaster} shows the results of this analysis for a medium size of the response batteries. 
Note that we still assume perfect knowledge of the response model.
We observe that, when we no longer assume
perfect knowledge of the system dynamics, the MCTS-based technique still attains significant improvements in the published price accuracy, reducing the price MAE of 12.8\% compared to the baseline. 
We further note that the minute-based quantiles of the prediction mostly decrease linearly as the quarter hour gets close to its end. However, different from the results obtained with perfect knowledge, we do not observe a significant reduction in price MAE in the earlier minutes of the quarter. This might be related to the higher uncertainties of the NRV forecasts: even though the tree can look forward in the quarter hour, its assumed future is inaccurate, as the forecasted NRV differs from the actual one.

\begin{figure}
    \centering
    \includegraphics[width=0.5\textwidth]{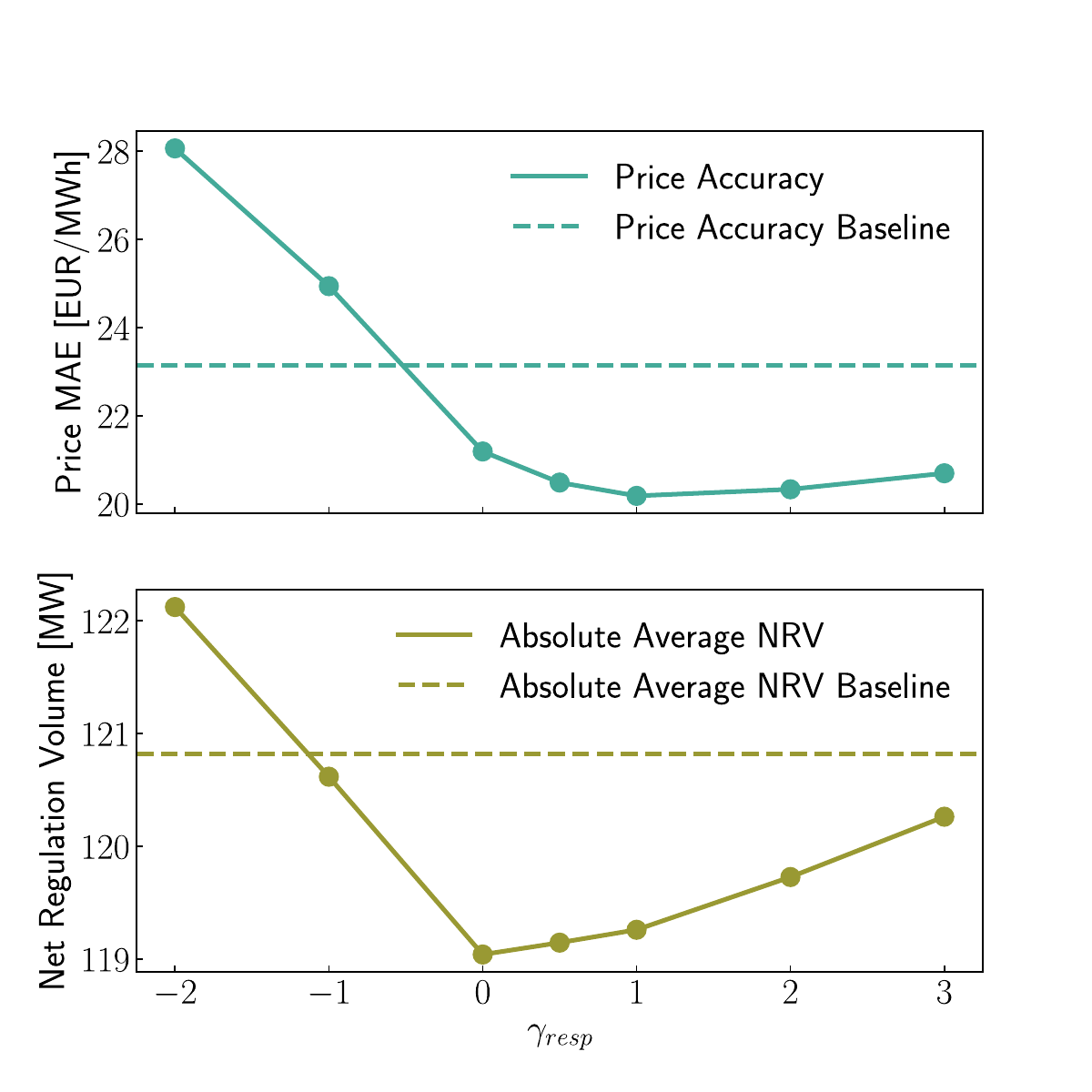}
    \caption{MCTS accuracy obtained when adding inaccuracies to the expected implicit response. The value of $\responseInaccuracy$ describes the discrepancy of the expected response vs. the actual response considered in the simulations. In terms of price accuracy, an inaccurate prediction in the response does not significantly influence the price accuracy, as long as the responses sign are correctly predicted. If that is not the case, the publication accuracy drops significantly. The averaged absolute NRV also appears to be particularly sensitive to the sign prediction of the response.}
    \label{fig:noisy_response_figure}
\end{figure}

Besides NRV forecasts, also the (implicit) BRP responses will be uncertain in practice.
Thus, next we analyze the effect of errors in predicting such responses in the simulations used in the MCTS approach. 
Specifically, we multiply the actual response (from the assumed battery systems) by an inaccuracy factor $\responseInaccuracy \in \mathbb{R}$. 
Hence, $\responseInaccuracy=1$ corresponds to the scenario with perfect knowledge of the implicit response, $\responseInaccuracy>1$ implies overestimating the expected response, and $\responseInaccuracy < 1$ represents underestimation of the response. 
Note that if $\responseInaccuracy < 0$, the sign of the implicit response gets wrongly predicted, implying that the expected response behavior is significantly different from the real one. 
Using a medium-sized battery response and including the NRV forecaster (\cref{sec:forecaster}) in our simulations, we varied the value of $\responseInaccuracy$.
The reward function considered in these experiments is once again described by \cref{eq:reward_function_1}.
\Cref{fig:noisy_response_figure} presents the resulting analysis.
The price accuracy seems to remain decently invariant when $\responseInaccuracy > 0$, meaning that the technique can obtain good results even when predicted responses are over-estimated ($\responseInaccuracy > 1$) or under-estimated ($0 \leq \responseInaccuracy < 1$). On the other hand, when the implicit response sign is wrongly predicted (\eg when an implicit response is expected to inject energy into the grid, but a withdrawal occurs instead), the price MAE increases substantially. We can then conclude that to obtain accurate price publications, it is not crucial to have accurate predictions of the absolute value of the implicit responses, but rather focus on predicting the correct sign. 
A similar conclusion can be drawn in terms of assessing the NRV magnitude, where we again observe a substantial increment when $\responseInaccuracy < 0$.
To conclude, after testing our technique with more realistic conditions, the MCTS algorithm still achieves remarkable results.

%=================================
\section{Final Discussion}
%=================================

\subsection{Conclusions}
In this research work, we presented a pioneering analysis of an MCTS-based real-time imbalance price publication tool from a TSO perspective. The technique uses state-of-the-art machine-learning forecasters to predict the future states of the grid, considering also the published prices and the consequential implicit responses of BRPs to increase the price accuracy. 
We benchmarked the technique with the publication method currently used in practice by the Belgian TSO, obtaining a 20.4\% reduction of the MAE on predicted imbalance prices when considering ideal conditions (perfect knowledge of the grid dynamics). Moreover, we evaluated the technique in a more realistic setup (using the NRV forecaster and adding inaccuracies in the response model), still achieving remarkable reduction of prediction errors, achieving $-$12.8\% lower MAE.
Last, we also analyzed the technique's ability to add secondary objectives beyond only price prediction accuracy.
Specifically, we added an NRV reduction term and a balancing cost reduction term. Our analysis indicates that TSOs can leverage the published prices to reduce the aforementioned values. However, as expected, this reduces the price prediction errors. Our results suggest that TSOs should carefully assess the overall cost-effectiveness, since this loss of accuracy could outweigh the (relatively limited) gains in terms of NRV or balancing cost reduction.

To the best of our knowledge, this is the first time a real-time publication technique for imbalance prices is proposed in the literature. Yet, we acknowledge that the presented work is to be seen as a promising proof-of-concept rather than a fully elaborated practical solution.
Still, in light of the growing integration of renewable energy sources into electricity grids and the resulting significance of demand response frameworks, we believe the topic of our work is highly relevant to the ongoing electrification of modern societies.

\subsection{Strengths and Limitations}
When using MCTS-based techniques like the one presented in this work, a series of advantages and drawbacks have to be taken into consideration. 
First, tree search techniques are model-based, meaning that meaningful insight can be infused into the publication method. 
Moreover, such techniques only require the model to be simulable. This allows non-linear forecasters such as neural networks (NNs) --- also in complex architectures such as the ones considered in this work --- to be included in the model. This is especially important when the system dynamics are particularly difficult to model with simpler tools, such as the case of the system imbalance (SI) dynamics of an electrical grid.
However, the obvious drawback of this characteristic is that a decently accurate model is required for the deployment of the technique. This increases the complexity of the task, as extra effort is required to create an accurate model. 

Another benefit of the technique is shown by the results obtained, as remarkable price accuracy has been achieved despite the technique being in its simpler form (\ie better results can be expected when the tree-search technique is expanded to more complex mechanics).
Despite the results obtained, a lack of explainability is a relevant limitation. Indeed, even though a certain grade of interpretability can be achieved through the tree structure, in case a series of price publications are particularly inaccurate (for example due to inaccuracies of the NRV forecaster), justifying these situations could be difficult.
Regarding computational costs, we note that the computational complexity is limited: all results presented were obtained using a modern laptop,\footnote{With a 12th Gen. Intel(R) Core(TM) i7-1265U--1.80\,GHz CPU and 16\,GB of RAM running on Python 3.10.11} on which a single forecast (to be generated every minute in practice) could be generated in $\sim$10 seconds.
When scaling the technique in real applications, a more advanced computational setup can increase the number of simulations involved in the tree search. Moreover, the technique inherently supports parallelization, which could significantly speed up 
the tree search process.
We can expect, by applying such refinements, to obtain even better results.

Last, we provide considerations regarding the evolution of the imbalance publication problem. From a game theory perspective, such a problem involves multiple parties trying to maximize their interests (namely the TSO, and the BRPs). A change in the TSO behavior (\eg a new publication technique) could cause a variation in the BRPs strategy. This feedback loop has to be regularly observed and considered to make sure that the technique is able to adapt to the evolution of the system. 

\subsection{Future Directions}
We conclude by suggesting different research directions that can be undertaken following our work. As previously mentioned, several improvements are required to refine the proposed technique and make it mature for real-world applications. We now present those we consider most relevant.

To start, more advanced MCTS algorithms should be applied and analyzed.
Especially the addition of NNs in the tree-search process has been shown to be remarkably effective~\cite{silver2017mastering,silver2018general}. 
Given the high amount of uncertainties involved in the problem, modifications of the MCTS algorithm to specifically address them should be studied.
MCTS literature has significantly expanded in the last decade, and works such as~\cite{antonoglou2021planning,dam2023monte} could be used as a starting point for such analyses.
Moreover, a more effective model of the implicit responses has to be studied. To get more accurate responses, more advanced techniques and analyses should be involved using specific data available to each TSO. Finally, given the relevance of the topic, we suggest future works to consider predictions of the imbalance prices not only in the current quarter hour, but also in the upcoming ones. 

\section*{Declaration of Competing Interest}
The authors declare that they have no known competing financial interests or personal relationships that could have appeared to influence the work reported in this paper.

\section*{Acknowledgments and funding}
Part of the research leading to these results has received funding from Agentschap Innoveren \& Ondernemen (VLAIO) as part of the Strategic Basic Research (SBO) program under the InduFlexControl-2 project. We also thank the AI4E team for their precious feedback, and Elia for their technical assistance.

\section*{Declaration of generative AI and AI-assisted technologies in the writing process}
During the preparation of this work the author(s) used ChatGPT 3.5 in order to enhance the language quality of the manuscript. After using this tool/service, the author(s) reviewed and edited the content as needed and take(s) full responsibility for the content of the publication.

\printcredits

\bibliography{bibl}

\appendix

\section{Calculation of $\alphaImb$}
\label{app:alpha}
To further incentivize imbalance participation when the SI magnitude is high, the Belgian TSO introduced a correction addend ${\alphaImb}_{,\,t}$ in the price calculation at timestep $t$. The value is obtained as follows~\cite{eliatariffs}:
\begin{equation}
    {\alphaImb}_{,\,t} \doteq a + \frac{b}{1 + \exp^{\frac{c-x_t}{d}}}\text{cp}_t \;\; ,
\end{equation}
where:
\begin{itemize}
    \item $a \doteq 0 \frac{\text{€}}{\text{MWh}}$
    \item $b \doteq 200 \frac{\text{€}}{\text{MWh}}$
    \item $c \doteq 450 \text{MWh}$
    \item $d \doteq 65 \text{MWh}$
    \item $x_t$ is the sliding average of the SI of current and previous quarter hours.
    \item $\text{cp}_t$ in timestep $t$ is obtained as:
    
    \begin{equation}
    \text{cp}_t \doteq 
        \begin{cases}
            \begin{rcases}
                0 & ; \quad \text{if } \MIPrice_t > 400\\
                \frac{400 - \MIPrice_t}{200} & ; \quad \text{if } 400 \geq \MIPrice_t \geq 200 \;\;\;\; \\
                1 & ; \quad \text{if } \MIPrice_t < 200\\
            \end{rcases}
        % \end{equation}
        & ;\; \text{if } \SI_t \leq 0
        \\
        \\
        % \begin{equation}
            \begin{rcases}
                0 & ; \quad \text{if } \MDPrice_t < -200\\
                \frac{\MDPrice_t + 200}{200} & ; \quad \text{if } -200 \leq \MDPrice_t \leq 0 \;\;\, \\
                1 & ; \quad \text{if } \MDPrice_t > 0\\
            \end{rcases}
            & ;\; \text{if } \SI_t > 0
        \end{cases}
    \end{equation}
    
\end{itemize}

\section{Calculation of $\omega_t$}
\label{app:omega}
When building our reward function that aimed to increase the price prediction accuracy (\cref{eq:reward_function_1}), we wanted to add a factor that assigned higher values to the rewards close to the end of the quarter. As a general set of rules for the weight factors, we identified the following:
\begin{enumerate}
    \item $\omega_t \in [0, 1] \;\; ; \; \forall t \in \mathbb{N}$
    \item $\omega_t \xrightarrow{t \rightarrow \left(\tq\right)^+} y_1$
    \item $\omega_t \xrightarrow{t \rightarrow \left(\tQ\right)^-} y_2$
\end{enumerate}
with $0 \leq y_1 \leq y_2 \leq 1$. We opted for the following function:
\begin{equation}
    \omega_{t} \doteq \frac{a^{d\tau(t)}-b}{c} \;\;,
\end{equation}
with $\tau(t) \doteq t \mod 15$. $d \in \mathbb{R}^+$ and $a \in \mathbb{R}^+$ are parameters defining the function slopes, while $b$ and $c$ are fixed to satisfy the previously stated property:
\begin{itemize}
    \item $b \doteq \frac{y_1-y_0a^d}{y_1-y_0}$
    \item $c \doteq \frac{1-b}{y_0}$
\end{itemize}
In our experiments, we set $a \doteq 2$, $d \doteq 5$, $y_1 \doteq 0.5$, and $y_2 \doteq 1$.

\section{Discretization of the Action Space}
\label{app:discretization_action_space}
When using the MCTS algorithm described in the paper, we need a discrete action space to work with. Because the MDP has a continuous action space, we were forced to discretize it. To do so, we first considered the approximated price obtained by using \cref{eq:price_formula} given the current NRV value. Then, we considered the prices of the bid ladder (of the current regulation) that are the closest to the marginal balancing price. Finally, we considered the first (\ie the cheapest) balancing bid of the opposite activation regulation (the decremental regulation if the current NRV is positive, and vice-versa). The union of those prices composes the action space $\mathcal{A}_t$, which is then dependent on the timestep index (because of the relation with the current NRV and bid ladder).

\section{Full Result List}
\label{app:full_metric_list}

To provide the full context of the results of the techniques, we provide a list with extra measurements performed in our experiments. That is shown in \cref{tab:full_metric_results}. Each value is the average over each quarter hour considered in the 10 days of evaluation when assuming ideal conditions (perfect knowledge of the grid's dynamics).
\begin{table*}[t]
\centering
\captionof{table}{Full results obtained in our experiments described in \cref{sec:results_exp1}. Each value in the table is quarter-averaged among the 10 days of evaluations.}
    \begin{tabular}{l|rr|rr|rr|rr}

        \multicolumn{1}{c}{} &
        \multicolumn{2}{c}{\textbf{No Response: }} & 
        \multicolumn{2}{c}{\textbf{Small Battery: }} & 
        \multicolumn{2}{c}{\textbf{Medium Battery: }} &
        \multicolumn{2}{c}{\textbf{Big Battery: }}
        \\
        \toprule
        \textbf{Response Magnitude} &
        \textbf{MCTS} & \textbf{Baseline} & 
        \textbf{MCTS} & \textbf{Baseline} & 
        \textbf{MCTS} & \textbf{Baseline} & 
        \textbf{MCTS} & \textbf{Baseline}
        \\
        \midrule
        \textbf{MAE published price} & 18.68 & 22.47 & 18.83 & 23.54 & 18.62 & 23.14 & 18.05 & 22.69 \\
        \textbf{MSE published price} & 4105.28 & 3729.39 & 4327.02 & 3933.96 & 4183.93 & 3904.73 & 3770.57 & 4100.19 \\
        \textbf{Absolute NRV} & 129.81 & 129.81 & 123.68 & 124.77 & 119.51 & 120.82 & 119.02 & 120.03 \\
        \textbf{Squared NRV} & 33434.65 & 33434.65 & 31091.97 & 31320.25 & 29122.61 & 29464.99 & 27484.52 & 27782.06 \\
        \textbf{NRVs Variation} & 39.87 & 39.87 & 40.04 & 40.65 & 42.98 & 43.79 & 57.27 & 57.48 \\
        \textbf{BRPs Profit} & 0.0 & 0.0 & 168.4 & 157.16 & 300.69 & 255.72 & 432.59 & 411.54 \\
        \textbf{NRV Sign Switches} & 0.48 & 0.48 & 0.51 & 0.64 & 0.6 & 0.93 & 0.75 & 1.57 \\
        \textbf{Published Prices Variation} & 5.5 & 6.28 & 7.58 & 7.18 & 9.96 & 9.0 & 14.56 & 12.99 \\
        \textbf{Balancing Costs} & 3439.06 & 3439.06 & 3353.07 & 3375.07 & 3292.99 & 3327.06 & 3332.02 & 3347.83 \\
        \bottomrule
    \end{tabular}
\label{tab:full_metric_results}
\end{table*}
In the table, the values \textbf{NRV Variation} and \textbf{Price Variation} correspond to the variation between each minute-based NRV (and published price) with the previous ones. The value \textbf{BRPs Profit} represents the average revenue obtained by the imbalance participation of the BRPs in each quarter hour.

\end{document}